\documentclass[aps,prd,secnumarabic,amssymb, amsmath,nobibnotes,nofootinbib,10pt]{revtex4}
\usepackage{amsfonts,amsmath,url, color}
\usepackage{bm, bbm}
\usepackage{graphicx}
\usepackage{bigints}

\newcommand{\hx}{\hat{x}}

\newcommand{\be}{\begin{equation}}
\newcommand{\ee}{\end{equation}}
\newcommand{\bea}{\begin{eqnarray}}
\newcommand{\eea}{\end{eqnarray}}

\usepackage{color}

\begin{document}

\title{Signal propagation on $\kappa$-Minkowski spacetime and non-local two-point functions}

\author{Michele Arzano}
\email{michele.arzano@roma1.infn.it}
\affiliation{Dipartimento di Fisica ``E. Pancini", Universit\`a di Napoli Federico II, I-80125 Napoli, Italy\\}
\affiliation{Dipartimento di Fisica,  ``Sapienza" Universit\`a di Roma, P.le A. Moro 2, 00185 Roma, Italy\\}
\author{Luca Tiberio Consoli}
\email{consoli.1558491@studenti.uniroma1.it}
\affiliation{Dipartimento di Fisica,  ``Sapienza" Universit\`a di Roma, P.le A. Moro 2, 00185 Roma, Italy\\}

\begin{abstract}
We study the propagation of quantum fields on $\kappa$-Minkowsi spacetime. Starting from the non-commutative partition function for a free field written in momentum space we derive the Feynman propagator and analyze the non-trivial singularity structure determined by the group manifold geometry of momentum space. The additional contributions due to such singularity structure result in a deformed field propagation which can be alternatively described in terms of an ordinary field propagation determined by a source with a blurred spacetime profile. We show that the $\kappa$-deformed Feynman propagator can be written in terms of vacuum expectation values of a {\it commutative} non-local quantum field. For sub-Planckian modes the $\kappa$-deformed propagator corresponds to the vacuum expectation value of the time-ordered product of non-local field operators while for trans-Plankian modes this is replaced by the Hadamard two-point function, the vacuum expectation value of the {\it anti-commutator} of non-local field operators.
\end{abstract}

\maketitle

\section{Introduction}

A recurring theme in quantum gravity research in recent years has been that of {\it dimensional reduction} (see \cite{Carlip:2017eud} for a comprehensive and up-to-date review).  Evidence from a variety of approaches to the problem indeed suggests that the effective dimensionality of spacetime might decrease below the infrared value of four as we probe shorter and shorter scales \cite{Ambjorn:2005db,Carlip:2016qrb,Lauscher:2005qz,Horava:2009if,Benedetti:2008gu,Modesto:2008jz,Benedetti:2009ge,Calcagni:2010bj,Calcagni:2013vsa,Calcagni:2014cza,Padmanabhan:2015vma,Carlip:2015mra}, with the majority of results pointing to a two-dimensional effective spacetime dimension at the Planck scale.\\
The classical description of spacetime in terms of a smooth manifold is expected to become unreliable at very short distances, when quantum gravitational effects cannot be neglected, and thus a notion of dimensionality in quantum gravity should be based on ``dimensional estimators" \cite{Carlip:2017eud} which can be generalized to quantum geometries. An example of such estimator is the notion of {\it spectral dimension} associated to a diffusion process determined, via the heat equation, by a Laplacian operator. Such characterization of dimensionality has been widely used in the literature to explore the running of dimensionality in various quantum gravity settings. The definition of spectral dimension relies on the existence of a Laplacian operator which governs the diffusion process in the particular model of quantum spacetime considered. It turns out that the non-trivial UV features of such diffusion process can be generally modelled by a deformation of the ordinary special relativistic energy-momentum dispersion relation \cite{Sotiriou:2011aa}. The existence of such departure from the usual relativistic relation between energy and momentum raises the issue of wether the phenomenon of running dimensionality could signal a breaking of Lorentz symmetry as we probe spacetime at the shortest scales \cite{Amelino-Camelia:2013cfa}. Models of {\it deformed} relativistic kinematics based on curved momentum space (with curvature scale proportional to the Planck energy $E_p \sim 10^{28}eV$) provide a framework in which a running spacetime dimensionality can co-exist with the notion of relativistic symmetries albeit of a deformed kind. An example of such models is given by the $\kappa$-Minkowski non-commutative spacetime \cite{Majid:1994cy}. The dual momentum space to such non-commutative spacetime is a non-abelian Lie group \cite{Arzano:2014jfa} whose manifold structure is given by ``half" of de Sitter space \cite{KowalskiGlikman:2004tz}. The group manifold structure of momentum space reflects on the structure of the generators of relativistic symmetries, which are now described by a {\it quantum deformation} of the Poincar\'e algebra known as the $\kappa$-Poincar\'e algebra \cite{Lukierski:1991pn, Lukierski:1992dt}.\\
A significant shortcoming of adopting the notion of spectral dimension to gain insight on the short distance features of spacetime is that the definition of spectral dimension is based on a fictitious diffusion process on a Euclidean space \cite{Amelino-Camelia:2013gna,Amelino-Camelia:2016sru}. Various alternative characterizations of the dimensionality of (quantum) spacetime in terms of estimators of more direct physical significance have appeared in recent literature \cite{Amelino-Camelia:2013gna,Amelino-Camelia:2016sru,Nozari:2015iba,Husain:2013zda,Alkofer:2016utc,Arzano:2017mdp,Arzano:2017goa}. For example, in the case of $\kappa$-Minkowski space, one can resort to field theoretic tools to describe the interaction potential between sources to gain information about the effective dimensionality of spacetime in the UV \cite{Arzano:2017uuh}. The analysis presented in \cite{Arzano:2017uuh} suggests that the non-trivial features of the Green's function leading to an effective dimensional reduction in the UV can be understood in terms of {\it fuzziness} of the spacetime profile of the source induced by non-commutativity. In this work we take as a starting point this observation and embark on a systematic analysis of the construction and behaviour of the Feynman propagator in $\kappa$-Minkowski space in order to gain a more complete picture of signal propagation in such non-commutative framework.\\
On one side our analysis aims to shed light on the connection between the path integral approach first pioneered in \cite{AmelinoCamelia:2001fd} and the canonical approach explored in \cite{Arzano:2010jw} to the quantization of fields on $\kappa$-Minkowski space. The bridge between these two pictures will be established via a  description of non-commutative fields on $\kappa$-Minkowski space in terms of {\it non-local} fields on ordinary Minkowkski space \cite{Freidel:2006gc,KowalskiGlikman:2009zu}. The work we present also aims at placing on firmer grounds the fuzzy spacetime picture of $\kappa$-deformed field propagation sketched in \cite{Arzano:2017uuh} providing an in depth description of the spacetime properties of the $\kappa$-deformed Feynman propagator.\\ 
In the next Section we introduce the basics of $\kappa$-Minkowski non-commutative spacetime, its associated momentum space and deformed symmetries. Moreover we collect all the necessary tools to carry out the field theoretic analysis in the following Sections, namely the non-commutative differential calculus, the notion of Weyl map, $\star$-product and integration on non-commutative spaces. Our analysis starts in Section 3 where we derive the $\kappa$-deformed Feynman propagator from the momentum space counterpart of the non-commutative partition function, analyze how this propagates field perturbations and interpret the results in terms of effects of spacetime fuzziness on sources and on the propagation process. In the remaining part of Section 3 we study the spacetime profile of the deformed propagator function according to the various possible values of its argument. In Section 4 we turn to the connection between non-commutative fields on $\kappa$-Minkowski space and a non-local field theory on ordinary Minkowski space. In Section 5 we proceed to a canonical quantization of the non-local field theory derived in the previous Section, and observe that the $\kappa$-deformed propagator can be obtained as the vacuum expectation value of the time-ordered product of non-local fields for ``sub-planckian" momenta (with modulus lower than the deformation scale $\kappa$), while it coincides with the non-local  Hadamard two-point function for trans-planckian momenta. The concluding Section is devoted to a summary and discussion of the results obtained.

\section{The $\kappa$-deformed field theorist toolbox}

The $\kappa$-Minkowski spacetime \cite{Majid:1994cy} is described by the four dimensional Lie algebra
\begin{equation}\label{K-M lie algebra}
[\hat{x}^{0},\hat{x}^{i}]=\frac{i}{\kappa}\hat{x}^{i} , \ \ \ \ \ \ [\hat{x}^{i},\hat{x}^{j}]=0\,,\,\,\,\,\,\,\,\,\,i,j=1,2,3.
\end{equation}
It will be useful to work with the following five-dimensional matrix representation\footnote{With low indices $ \hat{x}^{0}=-\hat{x}_{0} $, $ \hat{x}^{i}=\hat{x}_{i} $ the $ \kappa $-Minkowski defining commutator reads: $ [\hat{x}_{i},\hat{x}_{0}]=\frac{i}{\kappa}\hat{x}_{i} $.} of the $\kappa$-Minkowski Lie algebra
\begin{equation}\label{k-m matrices}
\big(\hat{x}_{0}\big)_{a}^{ \ b}=\frac{i}{\kappa}\begin{pmatrix} \  0 & \ \textbf{0}^{T} & 1 \  \\ \ \textbf{0} & \underline{\textbf{0}} & \textbf{0} \ \\ \ 1 & \ \ \textbf{0}^{T} & 0 \ \end{pmatrix}, \ \ \ \ \  \big(\hat{x}_{i}\big)_{a}^{ \ b}=\frac{i}{\kappa}\begin{pmatrix} \ 0 & \textbf{e}_{i}^{T} & 0 \ \\ \ \textbf{e}_{i} & \underline{\textbf{0}} & \textbf{e}_{i} \ \\ \ 0 & -\textbf{e}_{i}^{T} & 0 \ \end{pmatrix} ,
\end{equation}
where the 3-vector $ \textbf{e}_{i} $ has entry 1 at the i'th position (i.e. $ \textbf{e}_{i}^{T}=(\delta_{i}^{1},\delta_{i}^{2},\delta_{i}^{3}) $ and $ T $ denotes transposition), the 3-vector $ \textbf{0} $ is the null vector and $ \underline{\textbf{0}} $ is the $3\times 3$ null matrix. 
Let us notice that the algebra \eqref{K-M lie algebra} can be seen as a subalgebra of the five dimensional Lorentz algebra $ \mathfrak{so}(4,1) $. Indeed, using the Iwasawa decomposition, the latter can be written as the following direct sum 
\begin{equation}
\mathfrak{so}(4,1)=\mathfrak{k}\oplus\mathfrak{a}\oplus\mathfrak{n} \ ,
\end{equation}
where $ \mathfrak{k} $ is the four-dimensional Lorentz algebra $ \mathfrak{so}(3,1) $, the algebra $ \mathfrak{a} $ is one-dimensional and $ \mathfrak{n} $ is a three-dimensional nilpotent algebra. A representation of the algebras $\mathfrak{a}$ and $ \mathfrak{n} $ is given by the $ 5\times 5 $ matrices\footnote{It is easy to verify that the spatial generators are nilpotent, i.e. $ [(\hat{x}_{i})]^{3}=0 $. } $ \mathsf{a}=i\kappa\big(\hat{x}_{0}\big)_{a}^{ \ b} $ and $ \mathsf{n}_{i}=i\kappa\big(\hat{x}_{i}\big)_{a}^{ \ b} $  respectively, so that they satisfy the $ \kappa $-Minkowski-like commutation relation
\begin{equation}
[\mathsf{a},\mathsf{n}_{i}]=\mathsf{n}_{i} \ ,
\end{equation}
with all other commutator being zero. \\

Non-commutative ``plane waves" are obtained by exponentiating the generators of \eqref{K-M lie algebra}. The dimensionful parameters appearing in the argument of the exponential are interpreted as $\kappa$-deformed momenta. Two important points should be stressed which radically distinguish the present non-commutative scenario with the usual commutative Minkowski spacetime. Since the non-commuting coordinates belong to a Lie algebra, plane waves will be elements of the corresponding Lie group and, accordingly, momenta will be coordinates on such group. Moreover, since $\kappa$-Minkowski coordinates are non-commuting objects, there will be inequivalent ordering prescription for defining a non-commutative plane wave. To fix the ideas we will focus on the ordering convention where the time coordinate $ \hat{x}^{0} $ appears to the right
\begin{equation}\label{right ordered group element}
g=e^{ik_{i}\hat{x}^{i}}e^{ik_{0}\hat{x}^{0}} \ .
\end{equation}
The group element $g$ in the five-dimensional matrix representation reads
\begin{equation}\label{matrix plane wave}
G_{a}^{ \ b}=\begin{pmatrix} \  \cosh(\frac{k_{0}}{\kappa})+e^{\frac{k_{0}}{\kappa}}\frac{\textbf{k}^{2}}{2\kappa^{2}} & -\frac{1}{\kappa}\textbf{k}^{T} & \sinh(\frac{k_{0}}{\kappa})+e^{\frac{k_{0}}{\kappa}}\frac{\textbf{k}^{2}}{2\kappa^{2}} \ \\ \ -e^{\frac{k_{0}}{\kappa}}\frac{1}{\kappa}\textbf{k} & \mathbb{I}  & -e^{\frac{k_{0}}{\kappa}}\frac{1}{\kappa}\textbf{k} \ \\ \ \sinh(\frac{k_{0}}{\kappa})-e^{\frac{k_{0}}{\kappa}}\frac{\textbf{k}^{2}}{2\kappa^{2}} &  \ \ \frac{1}{\kappa}\textbf{k}^{T} & \cosh(\frac{k_{0}}{\kappa})-e^{\frac{k_{0}}{\kappa}}\frac{\textbf{k}^{2}}{2\kappa^{2}} \ 
\end{pmatrix} \ ,
\end{equation}
where $ \mathbb{I} $ denotes the $ 3\times 3 $ identity matrix and $ \textbf{k}^{T}=(k_{1},k_{2},k_{3}) $. The choice of ordering \eqref{right ordered group element} is quite natural in view of the Iwasawa decompostion defined above. In fact the group element is nothing but the product of the two group elements $N=\exp(k_{i}\mathsf{n}_{i}/\kappa) $ and $ A=\exp(k_{0}\mathsf{a}/\kappa) $ obtained by exponentiating the algebras $ \mathfrak{n} $ and $\mathfrak{a}$ respectively. From the Iwasawa decomposition of the five-dimensional Lorentz group $SO(4,1)= SO(3,1) N A$ we can characterize the group generated by the $\kappa$-Minkowski Lie algebra as the quotient $SO(4,1) / SO(3,1) \sim NA$, also denoted in the literature as $AN(3)$ (for more details see \cite{Arzano:2014jfa,KowalskiGlikman:2004tz}). The real parameters $k_0$, $\textbf{k}$ appearing in \eqref{right ordered group element} are coordinates on the $AN(3)$ group momentum space known as ``horospherical" coordinates \cite{Arzano:2014jua}.

In order to describe the manifold structure of the $AN(3)$ Lie group we act with the generic element $ g\in AN(3) $, in its $ 5\times 5 $ matrix representation $ G $, on the spacelike vector $ (0,0,0,0,\kappa) $ of the five-dimensional Minkowski (momentum) space. Writing the resulting vector in terms of global coordinates as $ G\cdot(0,0,0,0,\kappa)=(P_{0},P_{1},P_{2},P_{3},P_{4}) $ one gets
\bea\label{bicrossproduct to classical}
P_{0}&=&\kappa\sinh(\frac{k_{0}}{\kappa})+e^{\frac{k_{0}}{\kappa}}\frac{\textbf{k}^{2}}{2\kappa}  \nonumber \\
P_{i}& = &e^{\frac{k_{0}}{\kappa}}k_{i}  \nonumber\\
P_{4}& = &\cosh(\frac{k_{0}}{\kappa})-e^{\frac{k_{0}}{\kappa}}\frac{\textbf{k}^{2}}{2\kappa} \ . 
\eea
These coordinates satisfy the constraints 
\be
 -P_{0}^{2}+P^{i}P_{i}+P_{4}^{2}=\kappa^{2}\,,\,\,\,\,\,\, P_{0}+P_{4}>0\,.
\ee 
The former relation is nothing but the equation defining four-dimensional de Sitter space $dS_{4}$ in an embedding five-dimensional Minkowski space. The inequality $P_{0}+P_{4}>0$  restricts us to ``half"\footnote{The other half of $ dS_{4} $, i.e. the one identified by the condition $ P_{0}+P_{4}<0 $, can be obtained by replacing the action of $ G $ with $ G\cdot \mathcal{N} $, where
\[\mathcal{N}=\begin{pmatrix} -1 & \ \ \ \textbf{0}^{T} & \ 0 \\ \ \textbf{0} & \ \mathbb{I} & \ \textbf{0} \\ \ 0 & \ \ \ \textbf{0}^{T} & -1 \end{pmatrix} \ .
\] \label{footnote N}} of $dS_{4}$.

\subsection{The $\kappa$-Poincar\'e algebra}

As illustrated above, at the momentum space level, the non-commutativity of spacetime leads to momenta which belong to a non-abelian Lie group rather than to a vector space as in ordinary relativistic kinematics. It is thus natural to expect that this basic structural shift will affect dramatically the ordinary notions of relativistic symmetries as described by the Poincar\'e group. To understand how these structures are affected let us consider the product of two ``right-ordered"  $AN(3)$ group elements $g = e^{i k_{i} \hx^{i}} e^{i k_0 \hx^{0}}$ and $h = e^{i l_{i} \hx^{i}} e^{i l_0 \hx^{0}}$. This can be written as
\be\label{eq:3.1a}
g h = e^{i k_{i} \hx^{i}} e^{i k_0 \hx^{0}} e^{i l_{i} \hx^{i}} e^{i l_0 \hx^{0}} = e^{i (k_i \oplus l_i) \hx^{i}} e^{i (k_0 \oplus l_0) \hx^0}\,,
\ee
where $k_i \oplus l_i = k_i + e^{-k_0/\kappa} l_i$ and $k_0 \oplus l_0 = k_0 + l_0$. The addition law $k_{\mu} \oplus l_{\mu}=(k_0 \oplus l_0, k_i \oplus l_i)$ is clearly non-abelian, i.e. $k_{\mu} \oplus l_{\mu} \neq l_{\mu} \oplus k_{\mu}$, since the ``momentum" Lie group is non-abelian. The addition law for momenta reflects the composition of conserved quantum numbers associated to translation generators. In particular the familiar addition of  momenta can be seen as a consequence of the Leibiniz rule for the action of translation generators on multi-particle states.  The non-abelian composition of momenta thus reflects a {\it deformed} action of space translation generators. In the language of Hopf algebras \cite{Fuchs:1992nq} this can be expressed in terms of a non-trivial {\it coproduct} for the spatial translation generators $K_i$ which act diagonally on right-ordered plane waves
\be\label{copB1}
\Delta K_i = K_i \otimes 1 + e^{-K_0/\kappa} \otimes K_i\,, 
\ee
while the time translation generator acts according to the usual Leibniz rule expressed by the trivial coproduct
\be\label{copB2}
\Delta K_0 = K_0 \otimes 1 + 1 \otimes K_0\,.
\ee
Notice that in the limit $\kappa\rightarrow \infty$ the coproduct \eqref{copB1} reduces to the trivial one. In a similar fashion the group inversion is reflected in a non-trivial {\it antipode} for the generators
\begin{align}\label{eq:3.1}
%\Delta K_0 = K_0 \otimes 1 + 1 \otimes K_0\,, \qquad \Delta K_a = K_a \otimes 1 + e^{-K_0/\kappa} \otimes K_a\,, \nonumber\\ 
S(K_0) = -K_0\,, \qquad S(K_i) = -e^{K_0/\kappa} K_i\,,
\end{align}
which determines the appropriate generalization of momentum subtraction operation $\ominus$ needed in order for the basic relation $k_{\mu} \oplus (\ominus k_{\mu}) =0$ to hold.\\

It is natural at this point to ask how these {\it deformations} of the action of translation generators affect the other generators of relativistic symmetries. In particular whether the generators of rotations and boosts also exhibit non-trivial coproducts and antipodes, and if the deformations affect the ordinary structure of the commutators of the Poincar\'e algebra. In general both structures will be deformed. These non-trivial structures are mathematically described by a ``quantum deformation" of the Poincar\'e algebra: the $\kappa$-Poincar\'e Hopf algebra introduced in \cite{Lukierski:1991pn}. It turns out that the Lorentz sector of the $\kappa$-Poincar\'e algebra is characterized by trivial coproducts and antipodes for the generators of rotations, while those of the boost generators are deformed
\begin{align}\label{eq:3.6}
\Delta M_i & = M_i \otimes 1 + 1 \otimes M_i\,, \qquad S(M_i) = -M_i\,, \nonumber\\ 
\Delta N_i & = N_i \otimes 1 + e^{-K_0/\kappa} \otimes N_i + \frac{1}{\kappa} \varepsilon_{ijk} K^j \otimes M^k\,, \nonumber\\ 
S(N_i) & = -e^{K_0/\kappa} N_i + \frac{1}{\kappa} \varepsilon_{ijk} e^{K_0/\kappa} K^j M^k\,.
\end{align}
Notice that setting $K_0 = K_i = 0$, i.e.\! restricting to the Lorentz algebra, we recover a trivial Hopf algebra structure. The particular realization of the $\kappa$-Poincar\'e Hopf algebra in terms of the generators $\{K_{\mu}, M_i, N_i\}$, i.e. with translation generators associated to the parametrization of the $AN(3)$ group in terms of horospherical coordinates, is known in the literature as the {\it bicrossproduct} basis of the $\kappa$-Poincar\'e algebra \cite{Majid:1994cy}. One of the characterizing features of the bicrossproduct basis is that, while translation generators behave as ordinary four-vectors under rotations, the commutators between boosts and translation generators are deformed 
\begin{align}\label{eq:3.5}
[K_0, N_i] &= -i N_i\,, \nonumber\\ 
[K_i, N_j] &= -i\delta_{ij} \left(\frac{\kappa}{2} \left(1 - e^{-2 K_0/\kappa}\right) + \frac{1}{2\kappa} K_i K^i\right) + \frac{i}{\kappa} K_i K_j\,.
\end{align}
It can be shown \cite{Bruno:2001mw} that the deformed commutator between boosts and spatial translation generators leads to finite boost transformations for which the modulus of the spatial momentum approaches the UV value of $\kappa$, rather than diverging, in the limit of infinite boost parameter. This behaviour is typical of models based on non-linear deformations of relativistic kinematics known as Doubly Special Relativity \cite{AmelinoCamelia:2000ge,AmelinoCamelia:2000mn,AmelinoCamelia:2010pd,KowalskiGlikman:2002we,Agostini:2002yd}, widely popular over the past twenty years as effective models of Planck-scale kinematics incorporating the Planck energy, in our case identified with the UV deformation parameter $\kappa$, as an observer independent energy scale \cite{Magueijo:2001cr,Hossenfelder:2012jw}.

A rather important point to stress is that different choices of coordinates on the $AN(3)$ manifold will lead, in general, to different coproducts and antipodes for the associated translation generators. For example the relations above can be used to derive the coproducts and antipodes for translation generators $P_{\mu}$ associated to the {\it embedding coordinates} defined in \eqref{bicrossproduct to classical} 
\begin{align}\label{eq:3.2}
\Delta(P_{0})&=P_{0}\otimes P_{+} +P_{+}^{-1}\otimes P_{0}+\frac{1}{\kappa}\sum_{i=1}^{3}P_{i}P_{+}^{-1}\otimes P_{i}\,, \nonumber\\ 
\Delta(P_{i})& =P_{i}\otimes P_{+}+1\otimes P_{i} \,, \nonumber\\ 
\Delta(P_{4})& =P_{4}\otimes P_{+}-P_{+}^{-1}\otimes P_{0}-\frac{1}{\kappa}\sum_{i=1}^{3}P_{i}P_{+}^{-1}\otimes P_{i}\nonumber\\
S(P_{0})& =-P_{0}+\frac{1}{\kappa}\textbf{P}^{2}P_{+}^{-1}=\kappa P_{+}^{-1}-P_{4}\,, \qquad S(P_{i})=-P_{i}P_{+}^{-1}\,, \qquad S(P_{4})=P_{4}\,,
\end{align}
where $ P_{+}\equiv \frac{P_{0}+P_{4}}{\kappa}$. The corresponding coproducts and antipodes for rotations and boosts will be now given by
\bea 
\Delta(M_{i}) & = & M_{i}\otimes 1+1\otimes M_{i}\\
\Delta(N_{i})& = & N_{i}\otimes 1+P_{+}^{-1}\otimes N_{i}+\frac{\epsilon_{ijk}}{\kappa}P_{j}P_{+}^{-1}\otimes M_{k}\label{coboost}\\
S(M_{i})& = &-M_{i}\\
S(N_{i})& = & -N_{i}P_{+}+\frac{\epsilon_{ijk}}{\kappa}P_{j}M_{k}\label{antiboost}\,.
\eea
These generators are known in the literature as the ``classical" basis \cite{Kosinski:1994br,Borowiec:2009vb} of the $\kappa$-Poincar\'e algebra since, unlike the bicrossproduct basis reviewed above, their commutators are just the ones of the {\it ordinary Poincar\'e algebra}. This also implies that the mass Casimir invariant naturally associated to the generators $P_{\mu}$ is just the ordinary one
\be\label{casclas}
\mathcal{C}(P)=P_{\mu} P^{\mu}\,.
\ee
In other words in such classical basis the non-trivial features due to symmetry deformation manifest only in the ``co-algebra" sector (i.e. in the coproducts and antipodes) leaving unmodified the familiar Lie algebra structure of relativistic symmetries.\\
Finally, let us notice that in terms of bicrossproduct generators the Casimir is no longer quadratic and takes the form
\be\label{kCasimir}
\mathcal{C}(K) = \mathcal{C}_{\kappa}(K) \left(1+\frac{\mathcal{C}_{\kappa}(K)}{4\kappa^2}\right)
\ee
where $\mathcal{C}_{\kappa}(K)$ is the $\kappa$-deformed Casimir invariant naturally associated with the bicrossproduct basis \cite{Lukierski:1994}
\be
\mathcal{C}_{\kappa}(K)=\bigg(2\kappa\sinh\big(\frac{K_{0}}{2\kappa}\big)\bigg)^{2}-\textbf{K}^{2}e^{K_{0}/\kappa} \  .
\ee
Such Casimir determines a modification of the energy-momentum dispersion relation governed by the UV scale $\kappa$ common to many models of departures from ordinary relativistic kinematics at the Planck scale. Possible signatures of such deformed dispersion relations in the highest energy astrophysical phenomena have been among the leading candidate scenarios for experimental manifestations of quantum gravity effects \cite{AmelinoCamelia:2008qg, AmelinoCamelia:2009pg}.\\

\subsection{Non-commutative calculus}
As in ordinary field theory one expects the Casimir invariant \eqref{casclas} to have a ``coordinate space" counterpart in terms of a non-commutative wave operator. This will be written in terms of non-commutative differential operators complying with the non-trivial structure of the spacetime commutator \eqref{K-M lie algebra} and of the symmetry generators. In this Section we introduce the differential calculus needed to define such operators (for further technical details we refer the reader to \cite{Giller:1996,Kosinski:1995}).\\ 
As it is well known in the literature \cite{Sitarz:1994rh}, it is impossible to construct a four-dimensional set of non-commutative differentials which are also covariant under the action of $\kappa$-Poincar\'e generators\footnote{For instance, the 4D differential calculus used in \cite{Agostini:2006nc} and defined by the commutators \[ [\hat{x}_{0},d\hat{x}_{i}]=-\frac{i}{\kappa}d\hat{x}_{i} \ , \ \ \ [\hat{x}_{0},d\hat{x}_{0}]=0\ , \ \ \  [\hat{x}_{i},d\hat{x}_{\mu}]=0 \ , \] is covariant w.r.t. the action of translations alone, but is not $ \kappa $-Lorentz covariant. }.  Rather one has to resort to a five-dimensional set of non-commutative differentials $\{d\hat{x}_{0},d\hat{x}_{1},d\hat{x}_{2},d\hat{x}_{3},d\hat{x}_{4}\}$ with the following commutation relations with the $\kappa$-Minkowski coordinates
\[ [\hat{x}_{0},d\hat{x}_{0}]=\frac{i}{\kappa}d\hat{x}_{4} \ , \ \ \ [\hat{x}_{0},d\hat{x}_{i}]=0 \ , \ \ \ [\hat{x}_{0},d\hat{x}_{4}]=\frac{i}{\kappa}d\hat{x}_{0} \ ,
\]
\begin{equation}\label{differential commutator}
[\hat{x}_{i},d\hat{x}_{0}]=\frac{i}{\kappa}d\hat{x}_{i} \ , \ \ \ [\hat{x}_{i},d\hat{x}_{j}]=\delta_{ij}\frac{i}{\kappa}(d\hat{x}_{0}-d\hat{x}_{4}) \ , \ \ \ [\hat{x}_{i},d\hat{x}_{4}]=\frac{i}{\kappa}d\hat{x}_{i} \ .
\end{equation}
It can be checked by taking the differential of both sides of \eqref{K-M lie algebra} that these commutators are consistent with the non-commutative structure of spacetime and that all Jacobi identities involving differentials and non-commuting coordinates are satisfied. The Lorentz covariance of such relations can be easily checked using the relations \cite{Majid:1994cy}
\begin{equation}\label{k-p action}
N_{i}\triangleright\hat{x}_{0}=i\hat{x}_{i}, \ \ \ N_{i}\triangleright\hat{x}_{j}=i\delta_{ij}\hat{x}_{0},
\end{equation}
and extending the action to the differentials algebra in a natural way as\footnote{We also assume that the differential $ d\hx_{4} $ is $ \kappa $-Poincar\'e invariant $ \mathcal{P}_{\kappa}\triangleright d\hx_{4}=0 $, where $ \mathcal{P}_{\kappa} $ is a generic element of the $ \kappa $-Poincar\'e algebra.}
\be
N_{i}\triangleright d\hat{x}_{\mu}=d(N_{i}\triangleright\hat{x}_{\mu}) \ ,  \ \ \ \ \ \ N_{i}\triangleright(\hat{x}_{\mu}d\hx_\nu)=(N_{i}^{(1)}\triangleright\hat{x}_{\mu})(d(N_{i}^{(2)}\triangleright\hat{x}_{\nu})) \ ,
\ee
where we have used Sweedler notation $ \Delta(N)=N^{(1)}\otimes N^{(2)}=\sum_{a} N^{(1)a}\otimes N^{(2)a} $ for the coproduct in \eqref{coboost}.\\

A differential on the algebra of functions over $ \kappa $-Minkowski spacetime can be defined as
\begin{equation}
d=id\hat{x}^{a}\hat{\partial}_{a} \ ,
\end{equation}
where the derivatives $ \hat{\partial}_{a} $ are determined by requiring that the Leibniz rule for the differential is satisfied, as we now show. Working in the bicrossproduct basis the explicit form of the $ \hat{\partial}_{a} $ can be derived by first noting that, from the commutator $ [\hat{x}_{\mu},d\hat{x}_{a}]=(\hat{x}_{\mu})^{b}_{ \ a} d\hat{x}_{b} $, follows the identity
\begin{equation}\label{matrix differential plane waves}
\hat{e}_{k}d\hat{x}^{a}\hat{e}_{\ominus k}=d\hat{x}^{b}G_{b}^{ \ a} \ ,
\end{equation}
where $ G_{b}^{ \ a} $ is the matrix representation (\ref{matrix plane wave}) of the right-ordered plane wave $ g=e^{i k_{i} \hx^{i}} e^{i k_0 \hx^{0}}\equiv\hat{e}_{k} $. Imposing then the Leibniz rule for the differential $d$ on the product $ \hat{e}_{k}\hat{e}_{q} $ we get
\begin{multline}\label{leibniz derivatives}
d(\hat{e}_{k}\hat{e}_{q})=(d\hat{e}_{k})\hat{e}_{q}+\hat{e}_{k}(d\hat{e}_{q})=(id\hat{x}^{a}\hat{\partial}_{a}\hat{e}_{k})\hat{e}_{q}+\hat{e}_{k}(id\hat{x}^{a}\hat{\partial}_{a}\hat{e}_{q})=i\big[(d\hat{x}^{a}\hat{\partial}_{a}\hat{e}_{k})\hat{e}_{q}+\hat{e}_{k}(d\hat{x}^{a}\hat{e}_{\ominus k}\hat{e}_{k}\hat{\partial}_{a}\hat{e}_{q})\big]=  \\
=id\hat{x}^{a}\big[(\hat{\partial}_{a}\hat{e}_{k})\hat{e}_{q}+(G^{ \ b}_{a} \hat{e}_{k})(\hat{\partial}_{b} \hat{e}_{q})\big] \ , \ \ \ \ \ \ \ \ \ \ \ \ \ \ \ \ \ \ \ \ \ \ \ \ \ \ \ \ \ \ \ \ \ \ \ \ \ \ \ \ \  \ \ \ \ \ \ \ \ \ \ \ \ \ \ \ \ \ \ \ \ \ \ \ 
\end{multline}
where in in the second term of the third equality we have introduced $ \hat{e}_{\ominus k}\hat{e}_{k}=1 $ and in the last equality we used the relation (\ref{matrix differential plane waves}). Accordingly, looking at the first and the last terms of (\ref{leibniz derivatives}), we find that, in order for the differential $ d=id\hat{x}^{a}\hat{\partial}_{a} $ to satisfy the Leibniz rule, the derivatives must have coproducts
\begin{equation}
\Delta(\hat{\partial}_{a})=\hat{\partial}_{a}\otimes 1+G_{a}^{\ b}\otimes\hat{\partial}_{b} \, .
\end{equation}
It turns out that these coproducts reproduce the ones for the classical basis generators $ P_{a} $ in \eqref{eq:3.2}, with the coproduct of the operator $ \hat{\partial}_{4} $ corresponding to the coproduct of $ (\kappa-P_{4}) $. Therefore we can identify non-commutative derivatives associated with the 5-dimensional covariant calculus with translation generators of the classical basis. The action of the derivatives on right-ordered plane waves is then
\begin{equation}\label{action on plane waves differential}
\hat{\partial}_{\mu}\hat{e}_{k}=P_{\mu}(k)\hat{e}_{k} \ , \ \ \ \hat{\partial}_{4}\hat{e}_{k}=\big(\kappa-P_{4}(k)\big)\hat{e}_{k} \ ,
\end{equation}
where the explicit form of the classical basis momenta $ P_{a}(k) $ in terms of the bicrossproduct momenta is given by (\ref{bicrossproduct to classical}). One also defines conjugate operators $\hat{\partial}_{\mu}^{\dagger}$ whose action on plane waves is given by\footnote{Here we have used the fact that the hermitian conjugate of a plane wave involves the antipode map $ S(p) $ on its momentum $ \hat{e}_{k}^{\dagger}=\hat{e}_{S(k)}\equiv\hat{e}_{\ominus k} $.}
\begin{equation}
\hat{\partial}_{\mu}^{\dagger}\hat{e}_{k}\equiv(\hat{\partial}_{\mu}\hat{e}_{k}^{\dagger})^{\dagger}=S\big(P(k)\big)_{\mu}\hat{e}_{k} \ , \ \ \ \hat{\partial}_{4}^{\dagger}=\hat{\partial}_{4} \ ,
\end{equation}
reflecting the fact that $ P\big(S(k)\big)_{a}=S\big(P(k)\big)_{a} $.  

From the action of the derivatives on $\hat{e}_{k}  $ one can straightforwardly derive the action of the operators $ \hat{\partial}_{a} $ on the generic function of non-commuting coordinates $\hat{f}(\hat{x})$. This can be done by resorting to the following Fourier expansion \cite{KowalskiGlikman:2009zu,Freidel:2007hk,Guedes:2013vi,Arzano:2014ppa} in terms of right-ordered non-commutative plane waves $ \hat{e}_{k} $ 
\begin{equation}\label{inverse fourier bicrossproduct}
\hat{f}(\hat{x})=\int d\mu(k)\tilde{f}_{r}(k)\hat{e}_{k}(\hat{x}) \, ,
\end{equation}
where the integration measure $ d\mu(k) $ is the Haar measure\footnote{It is a left invariant measure $ d\mu(pk)=d\mu(k) $, and it is worth noticing that in horospherical coordinates it is just the diffeomorphism invariant measure on $dS_{4}$ corresponding to the cosmological metric $ -dk_{0}^{2}+e^{2k_{0}/\kappa}dk_{i}^{2} $. } on $AN(3)$ 
\begin{equation}\label{measure bicrossproduct}
d\mu(k)=\frac{e^{3k_{0}/\kappa}}{(2\pi)^{4}}dk_{0}d\textbf{k} \ ,
\end{equation}
which can be also expressed in terms of the ordinary Lebesgue measure on the five-dimensional embedding space $d^{5}{P}$ as
\be
d\mu(P)=\kappa\frac{\delta(P_{a}P^{a}-\kappa^{2})\theta(P_{0}+P_{4})}{(2\pi)^{4}}d^{5}{P} \, ,
\ee
and the subscript of $\tilde{f}_{r}(k)$ denotes that the Fourier transform is defined in terms of right-ordered non-commutative plane waves.

\subsection{Weyl maps and $\star$-product}
\label{wmsp}

The Weyl map is a useful tool first introduced in quantum mechanics to map classical observables (commuting functions on phase space) to quantum observables (functions of non-commuting operators). Due to ordering ambiguities on the non-commutative side Weyl maps are obviously not unique. In our context a Weyl map will map a function on commutative Minkowski space to a (suitably ordered) function on the non-commutative $\kappa$-Minkowski space.  

Let us focus on plane waves. As we will see, in our context the ordering ambiguity will reflect different choices of {\it bases} for the $\kappa$-deformed translation generators. We define the ``right-ordered" Weyl map $ \Omega_{r} $ as
\begin{equation}
\Omega_{r}(e^{ikx})=e^{ik_{i}\hat{x}^{i}}e^{ik_{0}\hat{x}^{0}}=\hat{e}_{k} \ , \ \ \ \  \Omega_{r}^{-1}(\hat{e}_{k})=e^{ikx} \,,
\end{equation}
i.e. an ordinary plane wave is mapped to an $AN(3)$ group element written in the decomposition \eqref{right ordered group element} in which the non-abelian generator $\hat{x}^{0}$ is always to the right.
%as we will see $\Omega_{r}$ maps ordinary commuting plane waves onto non-commuting plane waves which are eigenfunction of the bicrossproduct translation generators and thus the momenta $k_{\mu}$.....\\
One can associate a commutative function $ f_{r}(x)$ to $ \hat{f}(\hat{x}) $ via this Weyl map using the Fourier expansion (\ref{inverse fourier bicrossproduct})
\begin{equation}\label{frx}
f_{r}(x)=\Omega_{r}^{-1}(\hat{f}(\hat{x}))=\int d\mu(k)\tilde{f}_{r}(k)\Omega_{r}^{-1}\big(\hat{e}_{k}(\hat{x})\big)=\int d\mu(k)\tilde{f}_{r}(k)e^{ikx} \ .
\end{equation}
Among the possible Weyl maps to functions on $\kappa$-Minkowski, a preferred choice, which we denote $\Omega_{c}$, is given by the map leading to non-commutative plane waves on which the derivatives $ \hat{\partial}_{\mu} $ of the non-commutative differential calculus have ``classical" action, i.e.
\begin{equation}\label{weyl cb derivatives}
\hat{\partial}_{\mu}\triangleright\Omega_{c}(e^{ipx})=\Omega_{c}(-i\partial_{\mu}e^{ipx})=p_{\mu}\Omega_{c}(e^{ipx}) \ .
\end{equation}

The Weyl map $\Omega_{c}$ is related to the classical basis coordinates $ P_{a} $ and, as it can be easily checked confronting the actions (\ref{action on plane waves differential}) and (\ref{weyl cb derivatives}), it has the following action on plane waves
\be
\Omega_{c}(e^{iPx})=\hat{e}_{k(P)} \ , \ \ \ \ \Omega_{c}^{-1}(\hat{e}_{k(P)})=e^{iPx} \ ,
\ee
%or, analogously,
%\begin{equation}\label{definition c weyl}
%\Omega_{c}(e^{iP(k)x})=\hat{e}_{k} \ , \ \ \ \ \Omega_{c}^{-1}(\hat{e}_{k})=e^{iP(k)x} \ ,
%\end{equation}
that is, $ \Omega_{c} $ maps a commutative plane wave labeled by $ P $ to a right-ordered non-commutative plane wave whose four-momentum is $ k_{\mu}(P) $, where $ k(P) $ is the inverse transformation of (\ref{bicrossproduct to classical}). Therefore, following \eqref{inverse fourier bicrossproduct}, a non-commutative function $\hat{f}(\hat{x})$ can be expressed as
\begin{equation}
\hat{f}(\hat{x})=\int d\mu(P)\tilde{f}_{c}(P)\Omega_{c}(e^{iPx}) \,,
\end{equation}
where $ \tilde{f}_{c}(P)=\tilde{f}_{r}(k(P)) $, and the commutative function $ f_{c}(x)$ associated to $ \hat{f}(\hat{x}) $ through the inverse classical basis Weyl map is given by
\begin{equation}
f_{c}(x)=\Omega_{c}^{-1}(\hat{f}(\hat{x}))=\int d\mu(P)\tilde{f}_{c}(P)e^{iPx} \ .
\end{equation}
Using such map we can finally introduce a suitable notion of integration on $\kappa$-Minkowski space as follows
\begin{equation}\label{integral notion}
\widehat{\int \ } \hat{f}\equiv\int d^{4}x \ \Omega_c^{-1}\big(\hat{f}(\hat{x})\big)=\int d^{4}x \ f_{c}(x) \ .
\end{equation}
On the space of commutative functions obtained via the action of $\Omega_{c}^{-1}$, the non-commutativity of $\kappa$-Minkowski space is reflected in a non-trivial $\star$-product which replaces the ordinary commutative pointwise product. The star product associated to the Weyl map $ \Omega_{c} $ is defined by the relation
\begin{equation}
\hat{f}(\hat{x})\hat{g}(\hat{x})=\Omega_{c}\big(f_{c}(x)\big)\Omega_{c}\big(g_{c}(x)\big)=\Omega_{c}\big(f_{c}(x)\star g_{c}(x)\big) \ .
\end{equation}
It can be shown \cite{KowalskiGlikman:2009zu} that the explicit formula for products of the form $ f_{c}^{\dagger}\star g_{c}=\Omega_{c}^{-1}\big(\hat{f}^{\dagger}\hat{g}\big) $ has the rather simple expression
\begin{equation}\label{star product classical}
f_{c}^{\dagger}(x)\star g_{c}(x)=f_{c}^{\ast}(x)\sqrt{1+\square/\kappa^{2}}   g_{c}(x) \ ,
\end{equation}
where with $ f_{c}^{\dagger} $ we denote the $ \kappa $-Minkowski hermitian conjugation involving the antipode, e.g. $ \big(e^{iPx}\big)^{\dagger}=e^{iS(P)x} $, while $ f_{c}^{\ast} $ is just the standard complex conjugation. The $\star$-product \eqref{star product classical} can be used to define the Fourier transform 
\begin{equation}\label{fourier transform star product}
\tilde{f}_{c}(P)=\widehat{\int \ }\big[\Omega_{c}(e^{iPx})\big]^{\dagger}\hat{f}(\hat{x})=\int d^{4}x \ \big(e^{iPx}\big)^{\dagger}\star f_{c}(x) \,,
\end{equation}
which, taking into account the explicit form of the integration measure and the $\star$-product, leads to the two fundamental relations
\be\label{explicit fourier transform}
\tilde{f}_{c}(P)=\frac{\vert P_{4}\vert}{\kappa}\int d^{4}x \ e^{-iPx}f_{c}(x) \ ,  
\ee
and 
\begin{equation}\label{fourier transform star explicited}
f_{c}(x)=\int \frac{d^{4}P \ \theta(P_{0}+P_{4})}{(2\pi)^{4}\vert P_{4}\vert/\kappa}\tilde{f}_{c}(P)e^{iPx} \ ,
\end{equation}
where $ P_{4}=\pm\sqrt{\kappa^{2}-P^{2}} $ and $ P^{2}=-P_{0}^{2}+\textbf{P}^{2} $. It is worth noticing that the operator $ \sqrt{1+\square/\kappa^{2}} $, coming from the $\star$-product, in momentum space is nothing but $ \vert P_{4}\vert/\kappa $, i.e. the same term appearing in the denominator of the integration measure; this will lead to important simplifications in what follows.

We conclude this introductory Section with some remarks on Lorentz invariance. As first noted in \cite{Freidel:2007hk} and successively elaborated in \cite{Arzano:2009ci}, the momentum space suffers from a subtle form of Lorentz symmetry breaking. Namely, for any negative energy mode the allowed range of rapidities is bounded above. As we discussed in the previous Sections, the bicrossproduct coordinates cover only half of de Sitter space identified by the condition $ \kappa P_{+}(k)\equiv P_{0}(k)+P_{4}(k)>0 $, so that the momentum space is not the whole de Sitter space $ dS_{4} $. In the classical basis this restriction explicitly breaks Lorentz invariance since it is not preserved by boosts (remember that $ P_{0}$ transform as the 0-th component of a Lorentz vector while $ P_{4} $ is a Lorentz scalar). Indeed it takes a boost with finite rapidity to bring a point out of the the region $ P_{0}+P_{4}>0 $.

A way to circumvent this problem is to take as momentum space the full de Sitter space quotiented by reflections $ P_{a}\rightarrow -P_{a} $. In fact by reflections the sector $ P_{0}+P_{4}>0 $ is sent to its complement. This space is called the {\it elliptic} de Sitter space $ dS_{4}/\mathbb{Z}_{2} $. Accordingly, one can change the defining condition from  $ P_{0}+P_{4}>0 $  to $ P_{4}>0 $, which is clearly Lorentz invariant, by considering, instead of the sector identified by the conditions $ P_{0}+P_{4}>0$ and $P_{4}<0 $, its image under reflection, i.e. the sector with  $ P_{0}+P_{4}<0$ and $P_{4}>0 $. The Fourier transform $ \tilde{f}_{c}(P) $, defined so far only in the region $ P_{+}>0 $, is now defined on the whole de Sitter momentum space, which is even under the $ \mathbb{Z}_{2} $ identification $ \tilde{f}_{c}(P_{a})=\tilde{f}_{c}(-P_{a}) $. This suggest that, in the classical basis, the Lorentz invariant measure on $ dS_{4}/\mathbb{Z}_{2} $ will be
\begin{equation}\label{measurecl}
d\mu(P)=2\kappa\frac{\delta(P_{a}P^{a}-\kappa^{2})\theta(P_{4})}{(2\pi)^{4}}d^{5}{P} \ .
\end{equation}
Solving the delta function with respect to $ P_{4} $, we have that a non-commutative function $\hat{f}(\hx)$ can be Fourier expanded as
\begin{equation}\label{nc field expansion c.b.}
\hat{f}(\hat{x})=\int \frac{d^{4}P \ \theta(\kappa^{2}-P^{2})}{(2\pi)^{4}\vert P_{4}\vert/\kappa}\tilde{f}_{c}(P)\Omega_{c}(e^{iPx}) \ ,
\end{equation}
where the Heaviside step function ensures that $ P_{4}=\sqrt{\kappa^{2}-P^{2}}\in\mathbb{R} $.

\section{Free $\kappa$-deformed quantum fields: the Feynman propagator}

\subsection{The $\kappa$-deformed free field partition function}
We now move to the study of $\kappa$-deformed quantum fields. The $ \kappa$-Poincar\'e invariant action of a free massive complex scalar field is given by
\begin{equation}\label{action k scalar field}
S_{free}[\hat{\phi},\hat{\phi}^{\dagger}]=\widehat{\int \ } \big[ \big(\hat{\partial}_{\mu}\hat{\phi}\big)^{\dagger}\big(\hat{\partial}^{\mu}\hat{\phi}\big)+m^{2}\hat{\phi}^{\dagger}\hat{\phi}  \big] \ .
\end{equation}
The derivatives $ \hat{\partial}_{\mu} $ are those of the 5D bicovariant and $ \kappa$-Poincar\'e covariant differential calculus illustrated in Section 2.
%The antipodes $ S(P) $ (sometimes we will denote them with $ \ominus p $) are related to the $ \kappa $-Minkowski hermitian conjugation, e.g. $ \big[\Omega_{c.b.}\big(e^{iPx}\big)\big]^{\dagger}=\Omega_{c.b.}\big(e^{iS(P)x}\big) $, while the co-products $ \Delta(P) $ induce a $ \kappa $-deformed addition law of momenta\footnote{It is worth notice that in the limit $ \kappa\rightarrow\infty $ the antipodes $ S(P)\rightarrow-P $ and the deformed addition law $ P\oplus Q\rightarrow P+Q $ reduce to the standard ones.}
%\[
%(P\oplus Q)_{0}=P_{0}Q_{+}+\frac{Q_{0}}{P_{+}}+\frac{1}{\kappa}\frac{P_{i}Q^{i}}{P_{+}} \ , 
%\]
%\begin{equation}\label{Deformed addition law}
%(P\oplus Q)_{i}=P_{i}Q_{+}+Q_{i} \ .  \ \ \ \ \ \ \ \ \ \ \ \ \ \
%\end{equation}
From the action (\ref{action k scalar field}), making use of the coproduct properties of the $ \hat{\partial}_{\mu} $'s (recall that $\Delta(\hat{\partial}_{\mu})=\Delta(P_{\mu})$), one obtains the following equation of motion
\be\label{eomnc1}
(\hat{\partial}_{\mu}\hat{\partial}^{\mu}+m^{2})\hat{\phi}(\hat{x})=0 \ ,
\ee
and an identical one for $\hat{\phi}^{\dagger}(\hat{x})$ thanks to the property $ (\hat{\partial}_{\mu}\hat{\partial}^{\mu})^{\dagger}=\hat{\partial}_{\mu}\hat{\partial}^{\mu} $, which reflects the fact that, in the classical basis, the antipodes satisfy the relation $ S(P)_{\mu}S(P)^{\mu}=P_{\mu}P^{\mu} $.
%\begin{equation}\label{equation of motion noncommutative}
%\ (\hat{\partial}_{\mu}\hat{\partial}^{\mu}+m^{2})\hat{\phi}^{\dagger}(\hat{x})=0 \ .
%\end{equation}
%This property of the derivatives $ \hat{\partial}_{\mu} $ just reflects the fact that, in the classical basis, the antipodes satisfy the relation $ S(P)_{\mu}S(P)^{\mu}=P_{\mu}P^{\mu} $. 
Considering the Fourier expansion (\ref{nc field expansion c.b.}), the free action (\ref{action k scalar field}) can be expressed in momentum space as 
\begin{equation}\label{action in momentum space}
S_{free}[\tilde{\phi},\tilde{\phi}^{\ast}]=\int \frac{d\bar{\mu}(p)}{(2\pi)^{4}}\tilde{\phi}^{\ast}(p)\big(p_{\mu}p^{\mu}+m^{2}\big)\tilde{\phi}(p) \ ,
\end{equation}
where the measure is $d\bar{\mu}(p)=d^{4}p\theta(\kappa^{2}-p^{2})\kappa/\vert p_{4}\vert$, and we denoted the classical basis momenta with $ p_{a} $ and the commutative functions $\tilde{\phi}_{c}(p)$ simply with $ \tilde{\phi}(p) $. In deriving this last expression we have also used the relation (\ref{weyl cb derivatives}) for the action of derivatives on plane waves and the following relation for integration on $ \kappa $-Minkowski
\begin{equation}\label{eq. cap 3}
\widehat{\int_{\hat{x}} \ } \big[\Omega_{c}(e^{ipx})\big]^{\dagger}\Omega_{c}(e^{iqx})=\int d^{4}x \ \big(e^{ip_{\mu}x^{\mu}}\big)^{\ast}\sqrt{1+\square/\kappa^{2}} \ \big(e^{iq_{\mu}x^{\mu}}\big)=(2\pi)^{4}\frac{\vert p_{4}\vert}{\kappa}\delta\big(p_{0}-q_{0}\big)\delta\big(\textbf{p}-\textbf{q}\big) \ .
\end{equation}

Looking at the expression of the action (\ref{action in momentum space}), the simplifications introduced by working with the classical basis of the $ \kappa $-Poincar\'e  algebra become evident. Indeed, as a result of the fact that in the classical basis the algebraic sector is undeformed, the Casimir $ \mathcal{C}=p_{\mu}p^{\mu} $ appearing in (\ref{action in momentum space}) is the standard one. Therefore, the momentum space free action differs from the ordinary one only for the integration measure $ d\bar{\mu}(p) $.\\ 

%If we had chosen to work, for instance, with the right ordered Weyl map $ \Omega_{r} $, and consequently with the bicrossproduct basis of the $ \kappa $-Poincaré Hopf algebra, we would also have to deal with a $ \kappa $-deformed Casimir $ \mathcal{C}=(1-\frac{\mathcal{C}_{\kappa}}{4\kappa^{2}})\mathcal{C}_{\kappa} $, where
%\begin{equation}
%\mathcal{C}_{\kappa}(k)=\bigg(2\kappa\sinh\big(\frac{k_{0}}{2\kappa}\big)\bigg)^{2}-\textbf{k}^{2}e^{k_{0}/\kappa} \  .
%\end{equation} 
The action in momentum space (\ref{action in momentum space}) can be used to write down the partition function of the theory. A partition function obtained from a momentum space action of a $ \kappa $-deformed field was first used in \cite{AmelinoCamelia:2001fd}. However, at that time, a full understanding of the momentum space related to the $ \kappa $-Poincar\'e Hopf algebra had not yet been reached. Specifically, the fact that the space of momenta is the four-dimesional elliptic de Sitter space was not taken into account and, consequently, the exact form of the momentum space integration measure was not given explicitly in the analysis of \cite{AmelinoCamelia:2001fd}. A more recent use of the $\kappa$-deformed partition function, which implemented the non-trivial geometric features of the momentum space, has appeared in \cite{Arzano:2017uuh}. This work presented a field theoretic approach to the study of the potential between two static point sources in a non-commutative space. The partition function adopted in \cite{Arzano:2017uuh} can be straightforwardly generalized to the complex scalar field case as 
\begin{equation}\label{partition function def. noncommutative}
\bar{Z}[J,J^{\dagger}]=\int \mathcal{D}[\phi]\mathcal{D}[\phi^{\dagger}] \ e^{iS_{free}[\hat{\phi},\hat{\phi}^{\dagger}]+i\widehat{\bigintssss \ }\big[\hat{\phi}^{\dagger}\hat{J}+\hat{J}^{\dagger}\hat{\phi}\big]} \ , 
\end{equation}
where the action $ S_{free}[\hat{\phi},\hat{\phi}^{\dagger}] $ is the $ \kappa$-Poincar\'e invariant action (\ref{action k scalar field}). We focus on the normalized partition function
\begin{equation}
Z[J,J^{\dagger}]= \dfrac{\bar{Z}[J,J^{\dagger}]}{\bar{Z}[0,0]} \,.
\end{equation} 
In order to bring $ Z[J,J^{\dagger}] $ into a well-suited expression for the manipulation needed to extract the Feynman propagator, we rewrite the partition function in momentum space. Indeed, since the momentum space is a commutative space, here it is possible to handle the functional calculus (which we will illustrate below) unambiguously. Making use of \eqref{nc field expansion c.b.}, (\ref{action in momentum space}) and (\ref{eq. cap 3}), one obtains from (\ref{partition function def. noncommutative})
\begin{equation}\label{kpartition function}
Z[\tilde{J},\tilde{J}^{\ast}]=\dfrac{1}{\bar{Z}[0,0]} \int \mathcal{D}[\tilde{\phi}]\mathcal{D}[\tilde{\phi}^{\ast}]  \ e^{i\bigintssss\frac{d\bar{\mu}(p)}{(2\pi)^{4}} \ \big[ \tilde{\phi}^{\ast}(p)(p_{\mu}p^{\mu}+m^{2})\tilde{\phi}(p)+ \tilde{\phi}^{\ast}(p)\tilde{J}(p)+ \tilde{J}^{\ast}(p)\tilde{\phi}(p)\big]   } \ .
\end{equation}
The functional integration can now be carried out as an ordinary Gaussian integral and, after simple manipulations, one finds that
\begin{equation}\label{partition function in momentum}
Z[\tilde{J},\tilde{J}^{\ast}]= \exp\bigg(i\int\frac{d\bar{\mu}(p)}{(2\pi)^{4}} \dfrac{\tilde{J}^{\ast}(p)\tilde{J}(p)}{-p^{2}-m^{2}+i\varepsilon} \ \bigg) \ .
\end{equation}
In this last expression we introduced the usual shift $ m^{2}\rightarrow m^{2}-i\varepsilon $ to render the integral well defined.\\

In order to derive the Feynman propagator from the partition function $Z[\tilde{J},\tilde{J}^{\ast}]$ we need an appropriate generalization of the functional derivatives to the deformed setting. In particular one has to take into account the $\kappa $-deformed coproduct structure of the translation generators in \eqref{eq:3.2}, which leads to the following non-abelian addition laws for momenta
\bea\label{Deformed addition law}
(p\oplus q)_{0}& = &p_{0}q_{+}+\frac{q_{0}}{p_{+}}+\frac{1}{\kappa}\frac{p_{i}q^{i}}{p_{+}} \\ 
(p\oplus q)_{i}& = &p_{i}q_{+}+q_{i} \ . 
\eea
This issue was first faced in \cite{AmelinoCamelia:2001fd} where however, as recalled above, the explicit form of the momentum space integration measure was not taken into account. Nonetheless, for an explicit definition of the functional derivatives such information is needed. Indeed, an important ingredient in the construction of functional calculus is a notion of delta function on the space of momenta. We will consider a delta function compatible with the non-trivial momentum space measure $d\bar{\mu}(p)$ \cite{Freidel:2013rra,Gubitosi:2015osv, Arzano:2015gda}, i.e. such that 
\begin{equation}
\int d\bar{\mu}(q)\, \delta(p,q)\, f(q)=f(p) \ .
\end{equation}
It can be easily checked that such delta function is given by
\begin{equation}\label{momentum delta function}
\delta(p,q)=\delta\big((\ominus p)\oplus q\big)=\frac{\vert p_{4}\vert}{\kappa}\delta(p-q) \ ,
\end{equation}
%since 
%\begin{equation}\label{delta function 1}
%\int d\bar{\mu}(q) \delta\big((\ominus p)\oplus q\big)f(q)=f(p) \ .
%\end{equation}
%Notice that the delta function $ \delta\big((\ominus p)\oplus q\big) $ is compatible with the integral of plane waves on $\kappa$-Minkowski space (\ref{eq. cap 3}) since $ \big[\Omega_{c}(e^{ipx})\big]^{\dagger}\Omega_{c}(e^{iqx})=\Omega_{c}\big(e^{i((\ominus p)\oplus q)x}\big) $. 
and thus it is proportional to an ordinary delta function $ \delta(p-q) $, and, in particular, it is symmetric under the exchange of momenta in the argument $ \delta\big((\ominus p)\oplus q\big)=\delta\big((\ominus q)\oplus p\big) $. Let us mention that the other possible choice of delta function  $\delta\big( p\oplus (\ominus q)\big)$ would have been less natural since it carries and additional multiplicative factor, indeed from
\begin{equation}
\delta\big( p\oplus (\ominus q)\big)=\vert p_{+}\vert^{3}\frac{\vert p_{4}\vert}{\kappa}\delta(p-q)=\vert p_{+}\vert^{3}\delta\big((\ominus p)\oplus q\big) ,
\end{equation}
it is easily seen that
\begin{equation}\label{delta function 2}
\int d\bar{\mu}(q) \delta\big( p\oplus (\ominus q)\big)f(q)=\vert p_{+}\vert^{3}f(p) \ .
\end{equation}
Notice that the two delta functions $ \delta\big((\ominus p)\oplus q\big) $ and $ \delta\big( p\oplus (\ominus q)\big) $ are related by the antipode transformation $ (p,q)\rightarrow(\ominus p,\ominus q) $, and thus the appearance of the factor $ \vert p_{+}\vert^{3} $ is related to the Jacobian of the antipode map $ \big\vert J\big\lbrace\frac{\partial \ominus p}{\partial p}\big\rbrace\big\vert=\big\vert p_{+}^{-1}\big\vert^{3} $.

With the choice of delta function \eqref{momentum delta function} we can now proceed as in \cite{AmelinoCamelia:2001fd}, though specializing the discussion to a complex field, and define the following $\kappa$-deformed functional derivatives
\begin{equation}\label{functional derivatives momentum}
\begin{split} & \dfrac{\delta Z[\tilde{J},\tilde{J}^{\ast}]}{\delta \tilde{J}(q)}=\lim_{\epsilon\rightarrow 0}\frac{1}{\epsilon}\big\lbrace Z[\tilde{J}(p)+\epsilon\delta\big((\ominus p)\oplus q\big),\tilde{J}^{\ast}(p)]-Z[\tilde{J},\tilde{J}^{\ast}]\big\rbrace \ , \\
 & \dfrac{\delta Z[\tilde{J},\tilde{J}^{\ast}]}{\delta \tilde{J}^{\ast}(q)}=\lim_{\epsilon\rightarrow 0}\frac{1}{\epsilon}\big\lbrace Z[\tilde{J}(p),\tilde{J}^{\ast}(p)+\epsilon\delta\big((\ominus p)\oplus (\ominus q)\big)]-Z[\tilde{J},\tilde{J}^{\ast}]\big\rbrace \ ,  \end{split} 
\end{equation}
which clearly reduce to the ordinary definitions in the limit $ \kappa\rightarrow\infty $, given that $ \ominus p\rightarrow-p $ and $ p\oplus q\rightarrow p+q $.
These will be employed in the next Section to obtain the Feynman propagator on $ \kappa $-Minkowski non-commutative space.

\subsection{The Feynman propagator}

Following \cite{AmelinoCamelia:2001fd}, we define the $\kappa$-deformed Feynman propagator in terms of the functional derivative of the partition function (\ref{partition function in momentum}) with respect to incoming and outgoing source functions
\begin{equation}\label{functional derivative propagator}
i\tilde{\Delta}^{\kappa}_{F}(p,q)=\bigg(i\dfrac{\delta}{\delta \tilde{J}^{\ast}(p)}\bigg)\bigg(-i\dfrac{\delta}{\delta \tilde{J}(q)}\bigg)Z[\tilde{J},\tilde{J}^{\ast}]\bigg\vert_{\tilde{J},\tilde{J}^{\ast}=0}.
\end{equation}
Taking into account the relations (\ref{functional derivatives momentum}) one obtains
\begin{equation}\label{Feynman fourier transform}
i\tilde{\Delta}^{\kappa}_{F}(p,q)=i(2\pi)^{4}\frac{\delta\big((\ominus q)\oplus(\ominus p)\big)}{-q^{2}-m^{2}+i\varepsilon}=i(2\pi)^{4}\frac{\vert q_{4}\vert}{\kappa}\frac{\delta\big(q-S(p)\big)}{-q^{2}-m^{2}+i\varepsilon} \ ,
\end{equation}
where in the last equality we have expanded the delta function as in (\ref{momentum delta function}). Through the inverse Fourier transform \eqref{nc field expansion c.b.} it is then possible to obtain the free scalar Feynman propagator on $ \kappa $-Minkowski non-commutative space
\begin{equation}\label{Feynman propagator noncommutative}
i\hat{\Delta}_{F}^{\kappa}(\hat{x},\hat{y})=i\int\frac{d\bar{\mu}(p)}{(2\pi)^{4}}\frac{\Omega_{c}(e^{ipx})\big[\Omega_{c}(e^{ipy})\big]^{\dagger}}{-p^{2}-m^{2}+i\varepsilon} \ .
\end{equation}

A point that deserves to be stressed, is that $i\hat{\Delta}^{\kappa}_{F}(\hat{x},\hat{y})$ is not symmetric under exchange of its arguments. This property of the Feynman propagator (\ref{Feynman propagator noncommutative}) originates from the fact that, in the $ \kappa $-deformed setting, the hermitian conjugate of a plane wave involves the antipode map $ S(p) $ on its momentum. Such spacetime asymmetry of $i\hat{\Delta}^{\kappa}_{F}(\hat{x},\hat{y})$, as we will see, although it may seem puzzling, does not lead to an actual physical asymmetry of the $ \kappa $-Minkowski field propagation. Nonetheless, the combination of non-commutative plane waves appearing in (\ref{Feynman propagator noncommutative}), which is the cause of this concern, makes sure that the Feynman propagator is a Green's function of the $ \kappa $-Klein Gordon equation \eqref{eomnc1}, as we now show. As a first step, we define the non-commutative delta function $ \hat{\delta}(\hat{x},\hat{y}) $ using the $ \kappa $-Minkowski Fourier transform and antitransform \eqref{fourier transform star product} and \eqref{nc field expansion c.b.}
\begin{equation}
\hat{f}(\hat{x})=\int  \frac{d\bar{\mu}(p)}{(2\pi)^{4}} \ \tilde{f}(p) \ \Omega_{c}(e^{ipx})=\int  \frac{d\bar{\mu}(p)}{(2\pi)^{4}} \ \bigg( \widehat{\int_{\hat{y}} \ } \ \big[\Omega_{c}(e^{ipy})\big]^{\dagger}\hat{f}(\hat{y})\bigg) \Omega_{c}(e^{ipx})\ .
\end{equation}
By requiring 
\be\label{Delta combination Fourier}
\hat{f}(\hat{x})=\widehat{\int_{\hat{y}} \ }\hat{\delta}(\hat{x},\hat{y})\hat{f}(\hat{y})\ ,
\ee
we are led to define the non-commutative $\delta$-function\footnote{The non-commutative delta function $\hat{\delta}(\hat{x},\hat{y})$ also satisfies:\[ \widehat{\int_{\hat{x},\hat{y}}}\hat{\delta}(\hat{x},\hat{y})=1  \ . \]} 
\begin{equation}\label{Delta noncommutative}
\hat{\delta}(\hat{x},\hat{y})=\int  \frac{d\bar{\mu}(p)}{(2\pi)^{4}}\Omega_{c}(e^{ipx})\big[\Omega_{c}(e^{ipy})\big]^{\dagger} \ .
\end{equation}
Applying the $ \kappa $-Klein-Gordon operator to the Feynman propagator (\ref{Feynman propagator noncommutative}), and taking into account the expression of the delta function (\ref{Delta noncommutative}), we thus get
\begin{equation}\label{K Green function eq}
(\hat{\partial}_{\mu}\hat{\partial}^{\mu}+m^{2})\hat{\Delta}^{\kappa}_{F}(\hat{x},\hat{y})=-\hat{\delta}(\hat{x},\hat{y}) \ ,
\end{equation}
which shows that $ \hat{\Delta}_{F}^{\kappa}(\hat{x},\hat{y}) $ is a (non-commutative) Green's function for this operator. \\

We now get back to the issue of the spacetime asymmetry of the $ \kappa $-deformed propagator raised above and study how  $ \hat{\Delta}^{\kappa}_{F}(\hat{x},\hat{y}) $ propagates the field in the presence of a perturbation generated by an external source $ \hat{J}(\hat{x}) $. This exercise will also provide significant insight on the physical properties of the $ \kappa $-Minkowski Feynman propagator. Given the $ \kappa $-Klein-Gordon equation in the presence of a source
\begin{equation}
(\hat{\partial}_{\mu}\hat{\partial}^{\mu}+m^{2})\hat{\phi}(\hat{x})=\hat{J}(\hat{x}) \ ,
\end{equation}
using the equations (\ref{Delta combination Fourier}) and (\ref{K Green function eq}) we can write down the following solution
\begin{equation}\label{propagation law noncommutative}
\hat{\phi}(\hat{x})=-\widehat{\int_{\hat{y}} \ }\hat{\Delta}^{\kappa}_{F}(\hat{x},\hat{y})\hat{J}(\hat{y}) \ .
\end{equation}
In contrast with the standard commutative case, the $ \kappa $-Minkowski integral (\ref{propagation law noncommutative}) involves the product of two non-commutative functions and requires some additional care. As recalled in the previous Section, such integral can be defined via the ordinary Lebesgue integral introducing a non-commutaive $\star$-multiplication between the fields. Indeed, writing the source $ \hat{J}(\hat{y}) $ as a Fourier integral
\begin{equation}\label{J Fourier expansion}
\hat{J}(\hat{y})=\int \frac{d\bar{\mu}(p)}{(2\pi)^{4}} \ \tilde{J}(p) \ \Omega_{c}(e^{ipy}) \ ,
\end{equation}
we see that (\ref{propagation law noncommutative}) involves an integral of the form
\begin{equation}
\widehat{\int_{\hat{y}} \ } \big[\Omega_{c}(e^{ipy})\big]^{\dagger}\Omega_{c}(e^{iqy}) \ .
\end{equation}
This is just the integral in (\ref{eq. cap 3}) which, making use of the inverse Weyl map $ \Omega^{-1}_{c} $ and its associated star product, can be expressed as
\begin{equation}\label{Integral of the propagation law}
\widehat{\int_{\hat{y}} \ } \big[\Omega_{c}(e^{ipy})\big]^{\dagger}\Omega_{c}(e^{iqy})=\int d^{4}y \ \big(e^{iS(p)_{\mu}y^{\mu}}\big)\star\big(e^{iq_{\mu}y^{\mu}}\big)=\int d^{4}y \ e^{-ip_{\mu}y^{\mu}} \ \sqrt{1+\square/\kappa^{2}} \ e^{iq_{\mu}y^{\mu}} \ .
\end{equation}
We have therefore that the propagation law (\ref{propagation law noncommutative}) can be expressed in terms of commutative fields as
\begin{equation}\label{Propagation law commutative}
\phi(x)=-\int d^{4}y \ \Delta_{F}^{\kappa}(x-y) \ \sqrt{1+\square/\kappa^{2}} \ J(y) \ ,
\end{equation}
where $ \phi(x)=\Omega^{-1}_{c}\big(\hat{\phi}(\hat{x})\big) $, $ J(y)=\Omega^{-1}_{c}\big(\hat{J}(\hat{y})\big) $ and we defined the $ \kappa $-deformed Feynman propagator on {\it commutative} Minkowski space
\begin{equation}\label{kappa deformed feynman propagator}
i\Delta_{F}^{\kappa}(x-y)=i\int \frac{d\bar{\mu}(p)}{(2\pi)^{4}}\dfrac{e^{ip(x-y)}}{-p^{2}-m^{2}+i\varepsilon}\, .
\end{equation}
From \eqref{Propagation law commutative} we see that the the spacetime asimmetry of the $\kappa $-Minkowski Feynman propagator \eqref{Feynman propagator noncommutative} does not affect the actual propagation of the field. Indeed, such asymmetry is canceled by the star-product of (\ref{Integral of the propagation law}), leading to a field propagation governed by the spacetime symmetric $ \kappa $-deformed Feynman propagator (\ref{kappa deformed feynman propagator}).\\

Let us also notice that in the propagation law (\ref{Propagation law commutative}) we can make the star product term $ \sqrt{1+\square/\kappa^{2}} $ act either on the source $ J(y) $ or, equivalently\footnote{Given the propagation law $\phi(x)=-\int d^{4}y \ \Delta_{F}^{\kappa}(x-y) \ \sqrt{1+\square/\kappa^{2}} \ J(y)$
we can take the formal series expansion in powers of the d'Alembertian for the star product term $\phi(x)=-\int d^{4}y \ \sum_{n=0}^{\infty}a_{n}\Delta_{F}^{\kappa}(x-y) \   \square^{n} J(y)$ which, after integrating by parts becomes $\phi(x)=-\int d^{4}y \ \sum_{n=0}^{\infty} a_{n} \square^{n}\Delta_{F}^{\kappa}(x-y) \ J(y)$.}, on the $ \kappa $-deformed propagator $ \Delta_{F}^{\kappa}(x-y) $.

Acting on $ J(y) $ and noticing that in momentum space the term $ \sqrt{1+\square/\kappa^{2}} $ is equal to $ \vert p_{4}\vert/\kappa $, so that it cancels the same factor in the non-trivial integration measure $ d\bar{\mu}(p)$, one obtains
\begin{equation}\label{propagation law 1}
\phi(x)=-\int d^{4}y \ \Delta_{F}^{\kappa}(x-y) J_{cl}(y) \ ,
\end{equation}
where $ J_{cl}(y) $ is a {\it classical} source 
\begin{equation}\label{J standard Fourier expansion}
J_{cl}(y)=\int \frac{d^{4}p}{(2\pi)^{4}} \ \tilde{J}(p)\, e^{ipy} \ ,
\end{equation}
i.e. just an ordinary commutative function which, in particular, can describe a sharply localized source (e.g. a Dirac delta function).

\begin{figure}[h!]
\centering
\includegraphics[width=0.50 \textwidth] {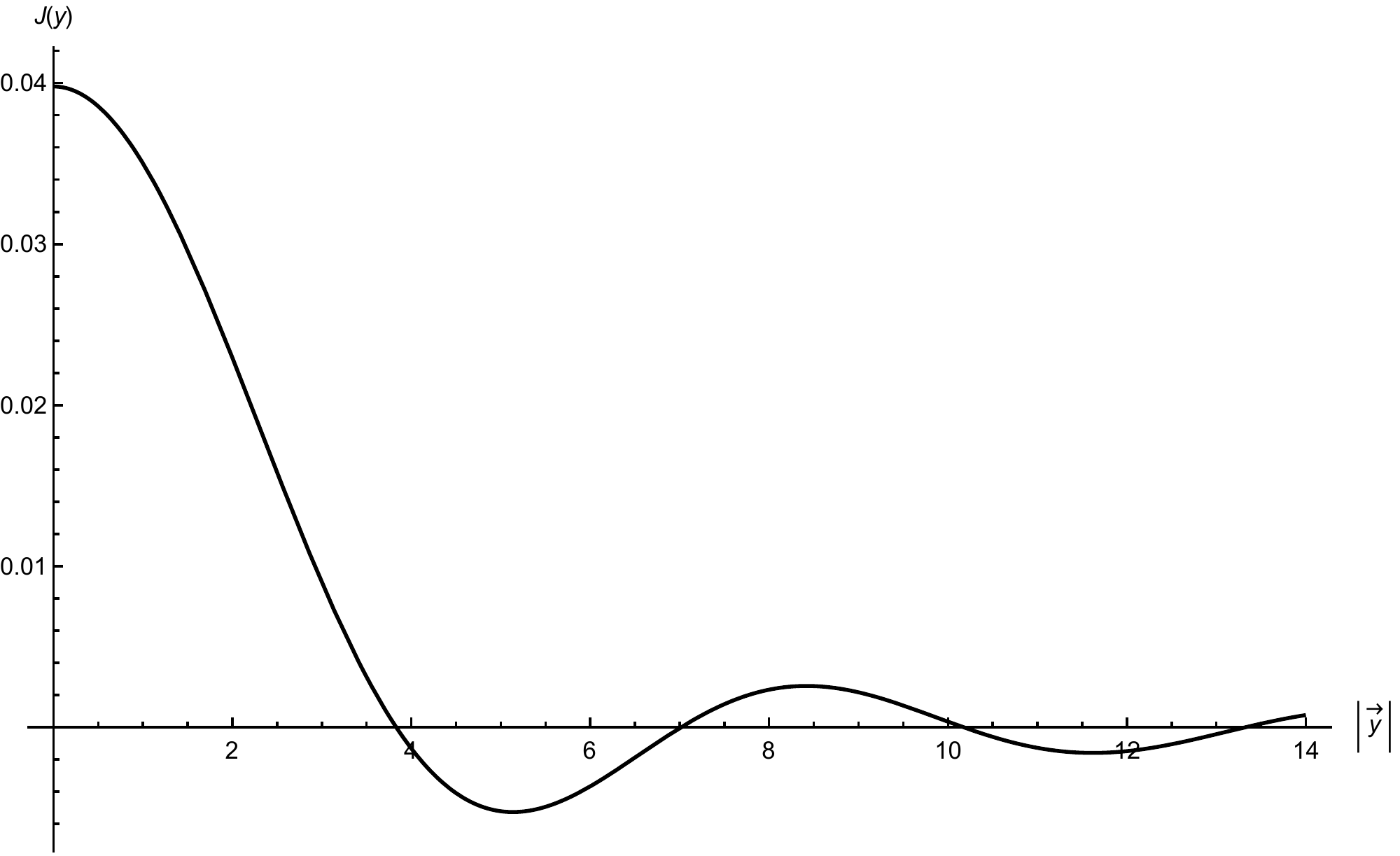}
\caption{{ {\footnotesize Smeared version $ J(y) $ of a classical source sharply localized in space $ J_{cl}(y)=\delta(\textbf{y}) $. Here we have set the deformation parameter $ \kappa=1 $ .}}} \label{smeared source}
\end{figure}

Acting instead with the star product term $ \sqrt{1+\square/\kappa^{2}} $ in \eqref{Propagation law commutative} on $ \Delta_{F}^{\kappa}(x-y) $, one gets
\begin{equation}\label{propagation law 2}
\phi(x)=-\int d^{4}y \ \Delta_{F}(x-y)J(y) \ ,
\end{equation}
where $ i\Delta_{F}(x-y) $ is the undeformed free scalar Feynman propagator
\be
i\Delta_{F}(x-y) =i\int\frac{d^{4}p}{(2\pi)^{4}} \ \frac{e^{ip(x-y)}}{-p^{2}-m^{2}+i\varepsilon}\,,
\ee
and where, in this case, the source function $ J(y) $ can not describe a point-like source due to the presence of the $ \kappa $-deformed integration measure $ d\bar{\mu}(p)= d^{4}p\, \theta(\kappa^{2}-p^{2})\kappa/\vert p_{4}\vert $ in its Fourier expansion (\ref{J Fourier expansion}). The source function $ J(y) $ can indeed be seen as a {\it smeared} version of the classical source (\ref{J standard Fourier expansion}). For instance, considering a classical source sharply localized in space $ J_{cl}(y)=\delta(\textbf{y}) $, for which $ \tilde{J}(p)=2\pi\delta(p_{0}) $, the source $ J(y) $ takes the form
\be
J(y)=2\int_{0}^\kappa\frac{dp \ p}{(2\pi)^2\sqrt{1-p^2/\kappa^2}}\frac{\sin(p\vert\textbf{y}\vert)}{\vert\textbf{y}\vert}=\frac{\kappa^2}{4\pi}\frac{J_{1}(\kappa\vert\textbf{y}\vert)}{\vert\textbf{y}\vert} \ ,
\ee
where $ J_1 $ is the Bessel function of the first kind (Fig. \ref{smeared source}).

Summarizing we have the following two pictures for the propagation of the non-commutative $\kappa$-scalar field in terms of commuting fields

\begin{enumerate}
\item Given a perturbation generated by a classical, and virtually sharply localized source $ J_{cl}(y) $, the field responds by propagating through the $ \kappa $-deformed Feynman propagator $ i\Delta_{F}^{\kappa}(x-y) $.

\item Given a perturbation generated by a $ \kappa $-deformed source $ J(y) $, the field responds by propagating through the standard Feynman propagator $ i\Delta_{F}(x-y) $.
\end{enumerate}

This result provides a concrete realization of the picture qualitatively outlined in \cite{Arzano:2017uuh}. There it was showed that for $\kappa$-deformed fields the Yukawa potential between two static point sources does not diverge in the short-distance limit, and that this feature could be interpreted in terms of point-like sources being effectively smoothed out by the UV features of the $ \kappa $-deformation. This is precisely what is realized in the propagation picture (\ref{propagation law 2}) outlined above. Although equally interesting, we postpone an analysis of such limitation in localizing sources in an arbitrarily small region to future studies, while, in this work, we will focus on the first picture of field propagation (\ref{propagation law 1}), where all the non-trivial structures due to the non-commutativity are contained in the $\kappa $-deformed Feynman propagator (\ref{kappa deformed feynman propagator}). The next step will be the explicit evaluation of the integral in \eqref{kappa deformed feynman propagator} to analyze the spacetime behaviour of $ i\Delta_{F}^{\kappa}(x) $.

\subsection{Spacetime profile of the $\kappa$-deformed propagator}

In order to study the spacetime properties of the $ \kappa $-deformed Feynman propagator, we must first stress that in the classical basis of the $\kappa$-Poincar\'e algebra the action of Lorentz transformations is {\it undeformed} (unlike for e.g, the bicrossproduct basis). We indeed have that the $ \kappa $-deformed Feynman propagator is manifestly Lorentz invariant
\begin{equation}\label{Feynman kappa spacetime}
i\Delta_{F}^{\kappa}(x-y)=i\int \frac{d^{4}p \ \theta(\kappa^{2}-p^{2})}{(2\pi)^{4}\sqrt{1-p^{2}/\kappa^{2}}}\dfrac{e^{ip(x-y)}}{-p^{2}-m^{2}+i\varepsilon} \ ,
\end{equation}
and thus the analysis of its spacetime behaviour can be divided, as in the standard case, according to whether the spacetime separation $ x^{\mu}-y^{\mu} $ is spacelike $ (x-y)^{2}>0 $, timelike $ (x-y)^{2}<0 $ or lightlike $ (x-y)^{2}=0 $. Let us notice, however, that the undeformed character of Lorentz transformations in the classical basis is limited to the one-particle sector of the theory. Indeed, when one considers multi-particle states the non-trivial co-algebra structure in \eqref{coboost}, \eqref{antiboost} enters the game, and the covariance of multiparticle states and observables should be assessed taking into account such deformed structures.

Before we start our analysis let us observe that the Feynman propagator (\ref{Feynman kappa spacetime}) can be conveniently expressed as
\begin{equation}\label{kappa deformed feynman expand 1}
i\Delta_{F}^{\kappa}(x-y)=i\int \frac{d^{3}p}{(2\pi)^{4}}e^{i\textbf{p}(\textbf{x}-\textbf{y})}\int\frac{dp_{0} \ \kappa \ \theta(\kappa^{2}-\textbf{p}^{2}+p_{0}^{2}) }{\sqrt{\kappa^{2}-\textbf{p}^{2}+p_{0}^{2}}}\frac{e^{-ip_{0}(x_{0}-y_{0})}}{p_{0}^{2}-\omega_{\textbf{p}}^{2}+i\varepsilon} \ ,
\end{equation}
where $ \omega_{\textbf{p}}=\sqrt{\textbf{p}^{2}+m^{2}} $. In order to carry out the $ p_{0} $ integral of this last expression we will make use, as in the standard case, of the Cauchy's residue theorem. However, we will have to deal, besides the non-trivial $ p_{0} $ range of integration, with a different singularity structure in the complex $ p_{0} $ plane, as compared to the standard case.

As a first step, in view of the condition $ \kappa^{2}-\textbf{p}^{2}+p_{0}^{2}>0 $, we split the range of integration in momentum space into two regions depending on whether $ \vert\textbf{p}\vert<\kappa $ or $ \vert\textbf{p}\vert>\kappa $ 
\begin{equation}\label{region a b}
\begin{split} A&=\big\lbrace \ \vert\textbf{p}\vert<\kappa \ \mid \ p_{0}\in \mathbb{R}  \ \big\rbrace \ , \\
B&=\big\lbrace \ \vert\textbf{p}\vert>\kappa \ \mid \ \vert p_{0}\vert> \sqrt{\textbf{p}^{2}-\kappa^{2}} \  \big\rbrace \ . \end{split}
\end{equation}
It is then possible to rewrite the expression (\ref{kappa deformed feynman expand 1}) as
\begin{equation}\label{kappa deformed feynman expand 2}
i\Delta_{F}^{\kappa}(x-y)=i\int_{\vert\textbf{p}\vert<\kappa} \frac{d^{3}p}{(2\pi)^{4}}e^{i\textbf{p}(\textbf{x}-\textbf{y})} \ \mathcal{I}^{A}_{\textbf{p}}(x_{0}-y_{0})+i\int_{\vert\textbf{p}\vert>\kappa} \frac{d^{3}p}{(2\pi)^{4}}e^{i\textbf{p}(\textbf{x}-\textbf{y})} \ \mathcal{I}^{B}_{\textbf{p}}(x_{0}-y_{0}) \ ,
\end{equation}
where we have defined the integrals
\[
\mathcal{I}^{A}_{\textbf{p}}(x_{0}-y_{0})=\int_{-\infty}^{+\infty}\frac{dp_{0} \ \kappa  }{\sqrt{p_{0}^{2}+\Omega_{A}^{2}}}\frac{e^{-ip_{0}(x_{0}-y_{0})}}{p_{0}^{2}-\omega_{\textbf{p}}^{2}+i\varepsilon} \ , \ \ \ \ \ \ \ \ \ \ \ \ \ \ \ \ \Omega_{A}=\sqrt{\kappa^{2}-\textbf{p}^{2}} \ ,   \ 
\]
\begin{equation}\label{integral IB}
\mathcal{I}^{B}_{\textbf{p}}(x_{0}-y_{0})=\int_{\Omega_{B}}^{\infty}\frac{dp_{0} \ \kappa  }{\sqrt{p_{0}^{2}-\Omega_{B}^{2}}}\frac{e^{-ip_{0}(x_{0}-y_{0})}+e^{ip_{0}(x_{0}-y_{0})}}{p_{0}^{2}-\omega_{\textbf{p}}^{2}+i\varepsilon} \ , \ \ \ \Omega_{B}=\sqrt{\textbf{p}^{2}-\kappa^{2}} \ .
\end{equation}
As we will see the singularity structure of $ i\Delta_{F}^{\kappa} $ in the complex $ p_{0} $ plane will differ in the sub-Planckian ($ \vert\textbf{p}\vert<\kappa $) and trans-Plankian ($ \vert\textbf{p}\vert>\kappa $) regions. Accordingly we will divide our analysis in two parts.

\subsubsection{Region $ A $: sub-Planckian modes} 

In the region $ A $ we deal with the integral 
\begin{equation}
\mathcal{I}_{\textbf{p}}^{A}(x_{0}-y_{0})=\bigintssss_{-\infty}^{+\infty} \dfrac{dp_{0} \ \kappa}{\sqrt{p_{0}^{2}+\Omega_{A}^{2}}} \dfrac{e^{-ip_{0}(x_{0}-y_{0})}}{(p_{0}-\omega_{\textbf{p}}+i\varepsilon)(p_{0}+\omega_{\textbf{p}}-i\varepsilon)} \ ,
\end{equation}
where $\Omega_{A}=\sqrt{\kappa^{2}-\textbf{p}^{2}}\in\mathbb{R}^{+} $ due to the fact that here $\vert\textbf{p}\vert<\kappa$. The square root at denominator can be written as $ \sqrt{(p_{0}-i\Omega_{A})(p_{0}+i\Omega_{A})} $, so that the points $ \pm i\Omega_{A} $ are two branch points of the square root type. As in the standard case one can solve the integral $ \mathcal{I}_{\textbf{p}}^{A}(x_{0}-y_{0}) $ employing Cauchy's residue theorem. However, it is first necessary to cut the complex $ p_{0} $ plane from $ +i\Omega_{A} $ to $ -i\Omega_{A} $ passing by infinity. Therefore, aside from the two undeformed simple poles in $ p_{0}=\pm\omega_{\textbf{p}}\mp i\varepsilon $, the singularity structure of the integrand in the complex $ p_{0} $ plane counts two branch cuts from $ +i\Omega_{A} $ to $ +i\infty $ and from $ -i\Omega_{A} $ to  $ -i\infty $. 

\begin{figure}[h!]
\centering
\includegraphics[width=0.45 \textwidth] {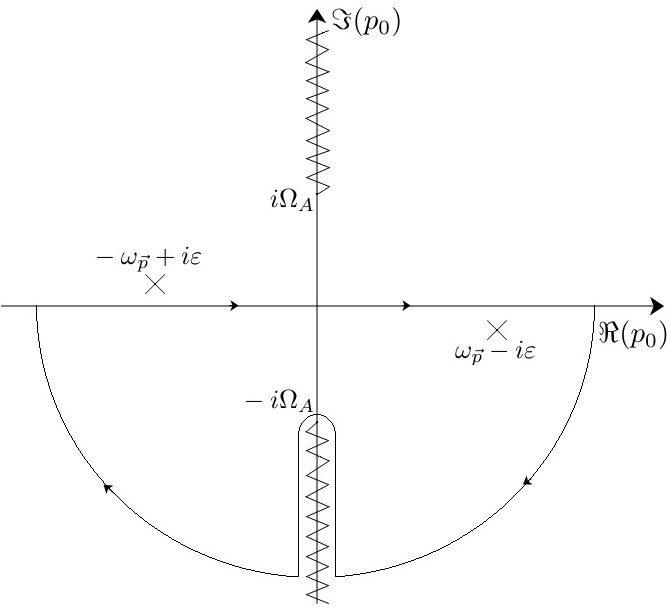}
\caption{{{\footnotesize Sector $ A $, path in the complex $ p_{0} $ plane used to evaluate $ \mathcal{I}_{\textbf{p}}^{A}(x_{0}-y_{0}) $ when $ x_{0}>y_{0} $.} }}\label{kpath p0 1}
\end{figure}

The path in the complex plane that we use to evaluate $ \mathcal{I}_{\textbf{p}}^{A}(x_{0}-y_{0}) $ when $ x_{0}>y_{0} $ is the one shown in Figure \ref{kpath p0 1}; while when $ x_{0}<y_{0} $ the path must be closed in the upper half plane. The result valid in both cases is
\begin{equation}\label{risultato A p0}
\mathcal{I}_{\textbf{p}}^{A}(x_{0}-y_{0})=\frac{-2\pi i}{\sqrt{1+m^{2}/\kappa^{2}}} \dfrac{e^{-i\omega_{\textbf{p}}\vert x_{0}-y_{0}\vert}}{2\omega_{\textbf{p}}}-2\int_{\Omega_{A}}^{\infty}\dfrac{dz \ \kappa}{\sqrt{z^{2}-\Omega_{A}^{2}}}\dfrac{e^{-z\vert x_{0}-y_{0}\vert}}{z^{2}+\omega_{\textbf{p}}^{2}} \ ,
\end{equation}
where the standard residue at the pole is only modified by the constant multiplicative factor $ (1+m^{2}/\kappa^{2})^{-1/2} $, while the new term comes from the discontinuity along the branch cut on the imaginary axis.

Let us pause for a moment on what consequences the presence of these imaginary axis branch cuts could have on the scalar field propagation. Let us consider, for simplicity, the case of a $ \kappa $-deformed retarded propagation of the field (for which both the simple poles are in the lower half plane) generated by a point like source $ J_{cl}(y)=\delta(y_{0}-t)\delta(\textbf{y}-\textbf{r}) $ which appears at $ y_{0}=t $ and suddenly disappears. The presence of the cuts on the imaginary axis makes it impossible to define a retarded propagator $ G_{R}^{ \kappa}(x-y) $ that vanishes for $ x_{0}<y_{0} $. As a result, the field will respond to the perturbation generated by $ J_{cl}(y) $ before the source itself switches on at $ y_{0}=t $. Although quite puzzling, this feature of the $ \kappa $-deformed retarded propagator can be interpreted as an effect of a spacetime fuzziness determined by the non-commutativity. Indeed, in the equivalent picture related to the field propagation (\ref{propagation law 2}), we would have a propagation mediated by the standard retarded propagator $ G_{R}(x-y) $, and a source $ J(y) $ that, being a smeared version of the point-like source $ J_{cl}(y) $, results active before the time $t $. Thus, in the first picture one has a $ \kappa $-deformed retarted propagator whose advanced effects are generated by the tachyon branch cuts, while, in the second picture, one has a source smoothed out by the $ \kappa $-deformation that, having support for times earlier than $ t $, allows a retarded field propagation before the time $ t $.

\subsubsection{Region $ B $: trans-Planckian modes}

\begin{figure}[h!]
\includegraphics[width=0.45 \textwidth] {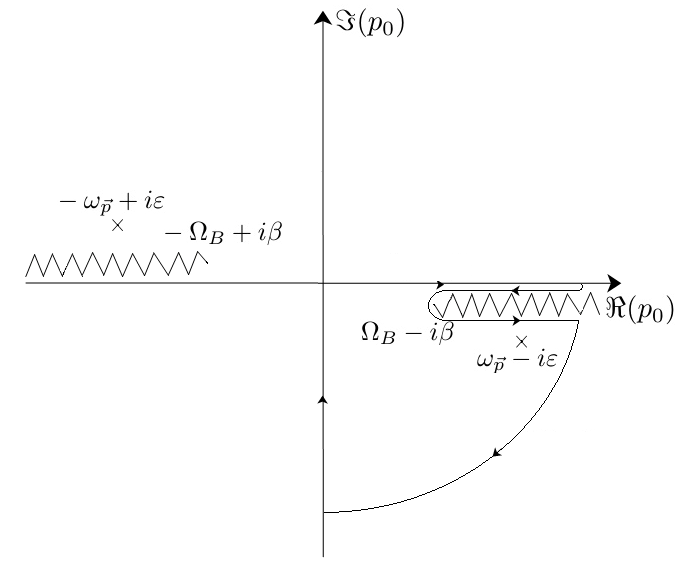}
\includegraphics[width=0.45 \textwidth] {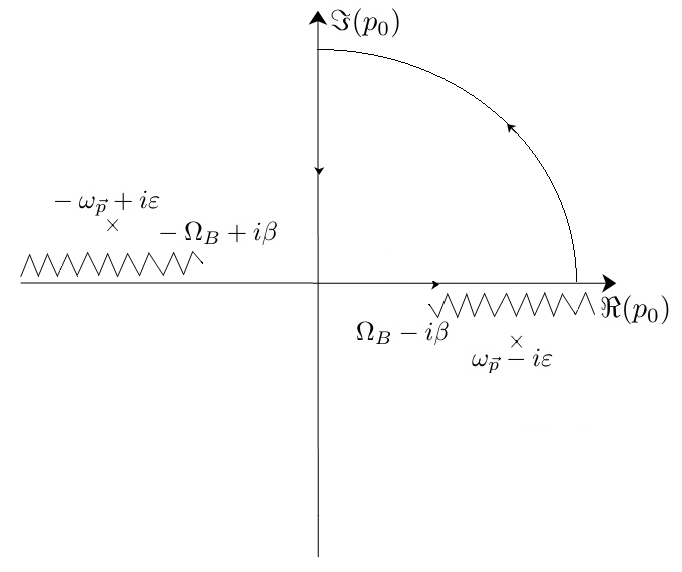}
\caption{{{\footnotesize Sector $ B $, paths in the complex $ p_{0} $ plane used to evaluate $ \mathcal{I}_{\textbf{p}}^{B(-)}(x_{0}-y_{0}) $  when $ x_{0}>y_{0} $ and $ x_{0}<y_{0} $ respectively. We have displaced the branch cuts by a small imaginary term in order to visually take into account the discontinuity along the cut.} }}\label{kpath p0 4}
\end{figure}

The contribution to $ i\Delta_{F}^{\kappa}(x-y) $ from the region $ B $ is given by the integral
\begin{equation}\label{parte b di k prop.}
\mathcal{I}_{\textbf{p}}^{B}(x_{0}-y_{0})=\int_{\Omega_{B}}^{+\infty} \dfrac{dp_{0} \ \kappa}{\sqrt{p_{0}^{2}-\Omega_{B}^{2}}} \dfrac{e^{-ip_{0}(x_{0}-y_{0})}+e^{ip_{0}(x_{0}-y_{0})}}{(p_{0}-\omega_{\textbf{p}}+i\varepsilon)(p_{0}+\omega_{\textbf{p}}-i\varepsilon)}=\mathcal{I}_{\textbf{p}}^{B(-)}(x_{0}-y_{0})+\mathcal{I}_{\textbf{p}}^{B(+)}(x_{0}-y_{0}) \ ,
\end{equation}
where we have denoted with $ \mathcal{I}_{\textbf{p}}^{B(\pm)} $ the positive and negative frequency part of $\mathcal{I}_{\textbf{p}}^{B}$. In this case $ \Omega_{B}=\sqrt{\textbf{p}^{2}-\kappa^{2}}\in\mathbb{R}^{+} $ due to the fact that $\vert\textbf{p}\vert>\kappa$. Therefore the real points $ \pm\Omega_{B} $ are branch points of the square root type and, in order to solve the integrals via Cauchy's residue theorem, we have to cut the complex $ p_{0} $ plane from $ \Omega_{B} $ to $ -\Omega_{B} $ passing by $ \infty $ (see Figure \ref{kpath p0 4}). We choose to focus on the negative frequency integral
\begin{equation}\label{pezzo di parte B di k prop.}
\mathcal{I}_{\textbf{p}}^{B(-)}(x_{0}-y_{0})=\int_{\Omega_{B}}^{+\infty} \dfrac{dp_{0} \ \kappa}{\sqrt{p_{0}^{2}-\Omega_{B}^{2}}} \dfrac{e^{-ip_{0}(x_{0}-y_{0})}}{(p_{0}-\omega_{\textbf{p}}+i\varepsilon)(p_{0}+\omega_{\textbf{p}}-i\varepsilon)} \ ,
\end{equation}
since it is then easy to extend the result to $\mathcal{I}_{\textbf{p}}^{B(+)}(x_{0}-y_{0}) $.

 The paths in the complex $ p_{0} $ plane that we use to evaluate $ \mathcal{I}_{\textbf{p}}^{B(-)} $ are shown in Figure \ref{kpath p0 4} and give as a result
\begin{equation}\label{parte B negative freq int}
\begin{split}\mathcal{I}_{\textbf{p}}^{B(-)}(x_{0}-y_{0})&=-\int_{0}^{\infty}\dfrac{dz \ \kappa}{\sqrt{\Omega_{B}^{2}+z^{2}}}\dfrac{e^{-z(x_{0}-y_{0})}}{z^{2}+\omega_{\textbf{p}}^{2}}-i\int_{0}^{\Omega_{B}}\dfrac{dp_{0} \ \kappa}{\sqrt{\Omega_{B}^{2}-p_{0}^{2}}}\dfrac{e^{-ip_{0}(x_{0}-y_{0})}}{p_{0}^{2}-\omega_{\textbf{p}}^{2}}+\frac{-2\pi i}{\sqrt{1+\frac{m^{2}}{\kappa^{2}}}}\dfrac{e^{-i\omega_{\textbf{p}}(x_{0}-y_{0})}}{2\omega_{\textbf{p}}} \ , \\
\mathcal{I}_{\textbf{p}}^{B(-)}(x_{0}-y_{0})&=-\int_{0}^{\infty}\dfrac{dz \ \kappa}{\sqrt{\Omega_{B}^{2}+z^{2}}}\dfrac{e^{z(x_{0}-y_{0})}}{z^{2}+\omega_{\textbf{p}}^{2}}+i\int_{0}^{\Omega_{B}}\dfrac{dp_{0} \ \kappa}{\sqrt{\Omega_{B}^{2}-p_{0}^{2}}}\dfrac{e^{-ip_{0}(x_{0}-y_{0})}}{p_{0}^{2}-\omega_{\textbf{p}}^{2}} \ , \end{split}
\end{equation}
for $ x_{0}>y_{0} $ and $ x_{0}<y_{0} $ respectively. From the formulas (\ref{parte B negative freq int}) the integral $ \mathcal{I}_{\textbf{p}}^{B(+)} $ can be straightforwardly evaluated. Then, considering the equation (\ref{parte b di k prop.}), one obtains for the $ p_{0} $ integral in the region $  B$ the expression
\begin{equation}\label{risultato B p0}
\mathcal{I}_{\textbf{p}}^{B}(x_{0}-y_{0})=\frac{-2\pi i}{\sqrt{1+m^{2}/\kappa^{2}}}\dfrac{e^{-i\omega_{\textbf{p}}\vert x_{0}-y_{0}\vert}}{2\omega_{\textbf{p}}}-2\int_{0}^{\Omega_{B}}\dfrac{dp_{0} \ \kappa}{\sqrt{\Omega_{B}^{2}-p_{0}^{2}}}\dfrac{\sin(p_{0}\vert x_{0}-y_{0}\vert)}{p_{0}^{2}-\omega_{\textbf{p}}^{2}}-2\int_{0}^{\infty}\dfrac{dz \ \kappa}{\sqrt{\Omega_{B}^{2}+z^{2}}}\dfrac{e^{-z\vert x_{0}-y_{0}\vert}}{z^{2}+\omega_{\textbf{p}}^{2}}. \\
\end{equation} 

Collecting now the results (\ref{risultato A p0}) and (\ref{risultato B p0}) obtained above, and plugging them in (\ref{kappa deformed feynman expand 2}), we can write down the final expression of the $ \kappa $-deformed Feynman propagator (\ref{kappa deformed feynman expand 1}) as
\begin{equation}\label{k propagatore esteso finale}
i\Delta^{\kappa}_{F}(x-y)=\frac{i\Delta_{F}(x-y)}{\sqrt{1+m^{2}/\kappa^{2}}} + i\Pi_{F}^{\kappa}(x-y) \ , 
\end{equation}
where $ i\Delta_{F}(x-y) $ is the standard free scalar Feynman propagator
\begin{equation}
i\Delta_{F}(x-y)=\int \dfrac{d^{3}p }{(2\pi)^{3} \ 2\omega_{\textbf{p}}}\ e^{i\textbf{p}\cdot(\textbf{x}-\textbf{y})}e^{-i\omega_{\textbf{p}}\vert x_{0}-y_{0}\vert} \ ,
\end{equation}
and $ i\Pi_{F}^{\kappa}(x-y) $ is given by
\begin{equation}
i\Pi_{F}^{\kappa}(x-y)=\int_{\vert\textbf{p}\vert<\kappa} \dfrac{d^{3}p}{(2\pi)^{4}}e^{i\textbf{p}\cdot(\textbf{x}-\textbf{y})} \ I^{A} \ +\int_{\vert\textbf{p}\vert>\kappa} \dfrac{d^{3}p}{(2\pi)^{4}}e^{i\textbf{p}\cdot(\textbf{x}-\textbf{y})}\big(I^{B}_{1}+I^{B}_{2}\big)  \,,
\end{equation}
where, in order to express $ i\Pi_{F}^{\kappa}(x-y) $ in a compact form, we have defined the integrals
\[
I^{A}=-2i\int_{\Omega_{A}}^{\infty}\dfrac{dz \ \kappa}{\sqrt{z^{2}-\Omega_{A}^{2}}}\dfrac{e^{-z\vert x_{0}-y_{0}\vert}}{z^{2}+\omega_{\textbf{p}}^{2}} \ ,  
\]
\begin{equation}\label{IA IB1 IB2}
I^{B}_{1}=-2i\int_{0}^{\infty}\dfrac{dz \ \kappa}{\sqrt{\Omega_{B}^{2}+z^{2}}}\dfrac{e^{-z\vert x_{0}-y_{0}\vert}}{z^{2}+\omega_{\textbf{p}}^{2}} \ , \ \ \ \ \ \ \ \ \ I^{B}_{2}=-2i\int_{0}^{\Omega_{B}}\dfrac{dp_{0} \ \kappa}{\sqrt{\Omega_{B}^{2}-p_{0}^{2}}}\dfrac{\sin(p_{0}\vert x_{0}-y_{0}\vert)}{p_{0}^{2}-\omega_{\textbf{p}}^{2}} \ .
\end{equation} 
Notice that in the limit $ \kappa\rightarrow\infty $ the $ \kappa $-deformed Feynman propagator reduces to the ordinary one, as it should be. Indeed, besides the trivial constant factor $ (1+m^{2}/\kappa^{2})^{-1/2} $ getting to 1, the term $ i\Pi_{F}^{\kappa}(x-y) $ disappears, since the integral $ I^{A} $ goes to zero, as $ \Omega_{A}\rightarrow\infty $, and the trans-Planckian sector $B$ disappears.  Notice also that the deformation term $ i\Pi_{F}^{\kappa}(x-y) $ is a pure imaginary quantity and thus the real part of $ i\Delta^{\kappa}_{F}(x-y) $ differs from the standard one only by the constant multiplicative factor\footnote{Which is, however, negligible if we identify $ \kappa $ with the Planck energy $ E_{p}\sim10^{28}eV $ and consider any of the particle masses of the standard model.} $ (1+m^{2}/\kappa^{2})^{-1/2} $.\\

Having derived the explicit expression for the $\kappa$-deformed propagator \eqref{k propagatore esteso finale},  we now proceed with the analysis of its spacetime behaviour. As customary we will divide the discussion in three cases depending on the sign of the norm of the spacetime separation $ x^{\mu} $.

\subsubsection{Lightlike separation}

When the spacetime separation $ x^{\mu} $ is a null vector, i.e. we are on the light cone $ x^{0}=\vert\textbf{x}\vert $, with a Lorentz transformation one can set $ x^{\mu}=(0,\textbf{0}) $. Therefore we consider the following $ \kappa $-deformed Feynman propagator 
\begin{equation}\label{k prop. lightlike}
i\Delta^{\kappa}_{F}(0)=\frac{i\Delta_{F}(0)}{\sqrt{1+m^{2}/\kappa^{2}}}+i\Pi_{F}^{\kappa}(0) \ . 
\end{equation}
The three integrals (\ref{IA IB1 IB2}) that define $ i\Pi_{F}^{\kappa}(0) $ can be explicitly computed as
\begin{equation}
I^{A}=-2i\dfrac{\sinh^{-1}\bigg(\frac{\omega_{\textbf{p}}}{\sqrt{\kappa^{2}-\textbf{p}^{2}}}\bigg)}{\sqrt{1+m^{2}/\kappa^{2}} \ \omega_{\textbf{p}}} \ , \ \ \ \ \ \ \ \ I^{B}_{1}=-2i\frac{\cosh^{-1}\bigg(\frac{\omega_{\textbf{p}}}{\sqrt{\textbf{p}^{2}-\kappa^{2}}} \bigg)}{\sqrt{1+m^{2}/\kappa^{2}} \ \omega_{\textbf{p}}}\ ,  \ \ \ \ \ \ \ \ I^{B}_{2}=0 \ ,
\end{equation}
so that, introducing spherical polar coordinates and performing the angular integration, one gets
\begin{equation}\label{Lightlike PI}
i\Pi_{F}^{\kappa}(0)= \frac{-4i}{\sqrt{1+m^{2}/\kappa^{2}}}\bigg[ \int_{0}^{\kappa}\frac{dp}{(2\pi)^{3}}\frac{p^{2}}{\omega_{p}} \sinh^{-1}\bigg(\frac{\omega_{p}}{\sqrt{\kappa^{2}-p^{2}}}\bigg)+\int_{\kappa}^{\infty}\frac{dp}{(2\pi)^{3}}\frac{p^{2}}{\omega_{p}}\cosh^{-1}\bigg(\frac{\omega_{p}}{\sqrt{p^{2}-\kappa^{2}}}\bigg)\bigg] \ ,
\end{equation}
where $ \vert\textbf{p}\vert=p $.

According to (\ref{k prop. lightlike}), on the light cone, the $ \kappa $-deformed Feynman propagator has the standard real quadratic divergence coming from $ i\Delta_{F}(0) $ and, in addition, due to the novel contribution of $ i\Pi_{F}^{\kappa}(0) $, it now exhibits also an imaginary divergent part. This imaginary contribution is due to the second integral in the square brackets of (\ref{Lightlike PI}). Indeed, given that 
\begin{equation}
\cosh^{-1}\bigg(\frac{\omega_{p}}{\sqrt{p^{2}-\kappa^{2}}}\bigg)\sim\frac{\sqrt{\kappa^{2}+m^{2}}}{p}+O\bigg(\frac{1}{p^{2}}\bigg) \ , \ \ \ \ \ p\rightarrow\infty \ ,
\end{equation}
the integrand $ \big[\frac{p^{2}}{\omega_{p}}\cosh^{-1}\big(\frac{\omega_{p}}{\sqrt{p^{2}-\kappa^{2}}}\big)\big] $, for large values of $ p $, approaches a constant so that the imaginary contribution of $ i\Pi_{F}^{\kappa}(0) $ is linearly divergent.

\subsubsection{Spacelike separation}

For spacelike separation $ x^{2}>0 $, one can set $ x^{0}=0 $ and $ \vert\textbf{x}\vert=\sqrt{x^{2}}=r $. Taking into account the equation (\ref{k propagatore esteso finale}), the $ \kappa $-deformed Feynman propagator reads
\begin{equation}\label{k prop. spacelike}
i\Delta^{\kappa}_{F}(r)=\frac{i\Delta_{F}(r)}{\sqrt{1+m^{2}/\kappa^{2}}}+i\Pi_{F}^{\kappa}(r) \ .
\end{equation}

\begin{figure}[h!]
\includegraphics[width=0.45 \textwidth] {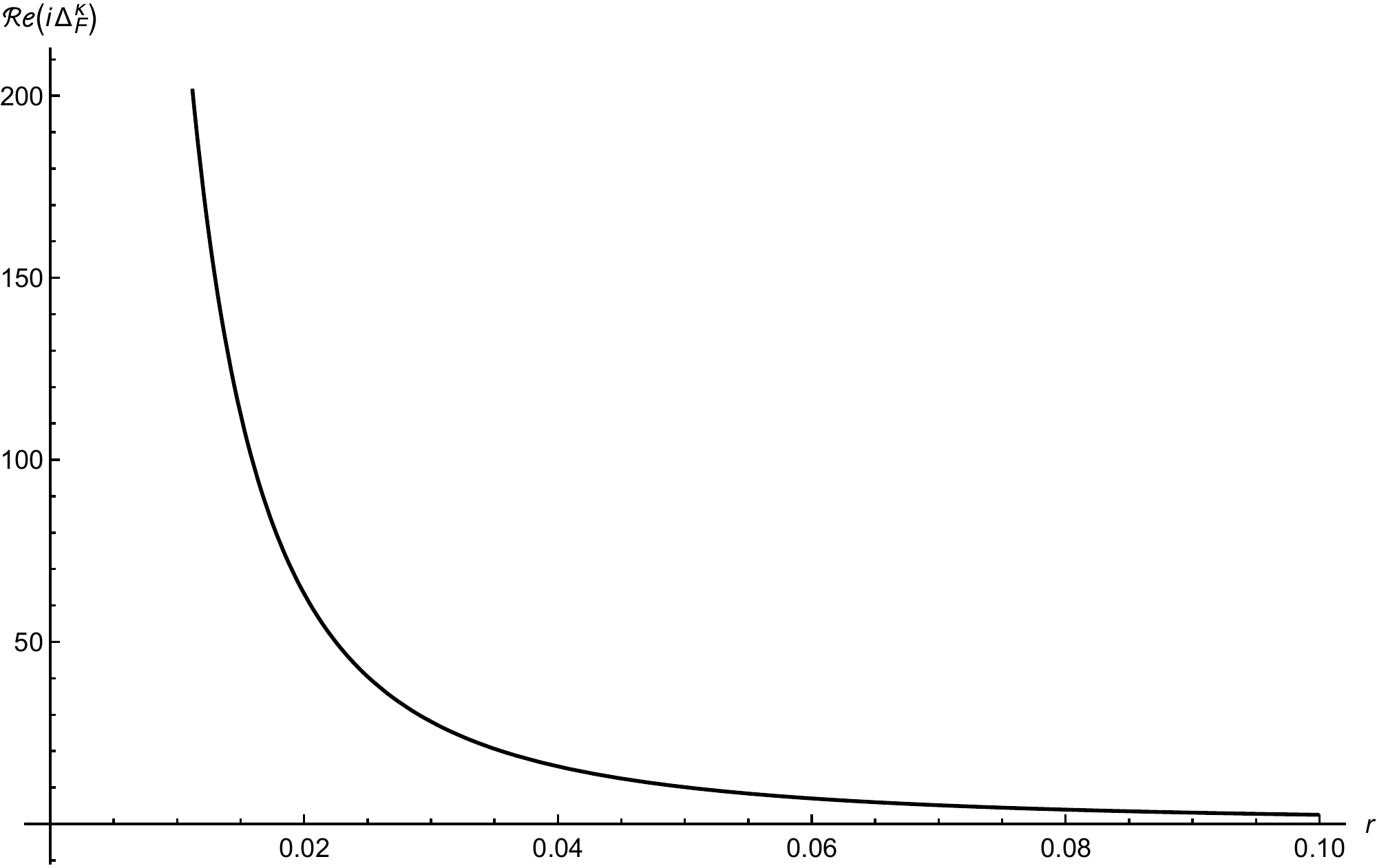}
\includegraphics[width=0.45 \textwidth] {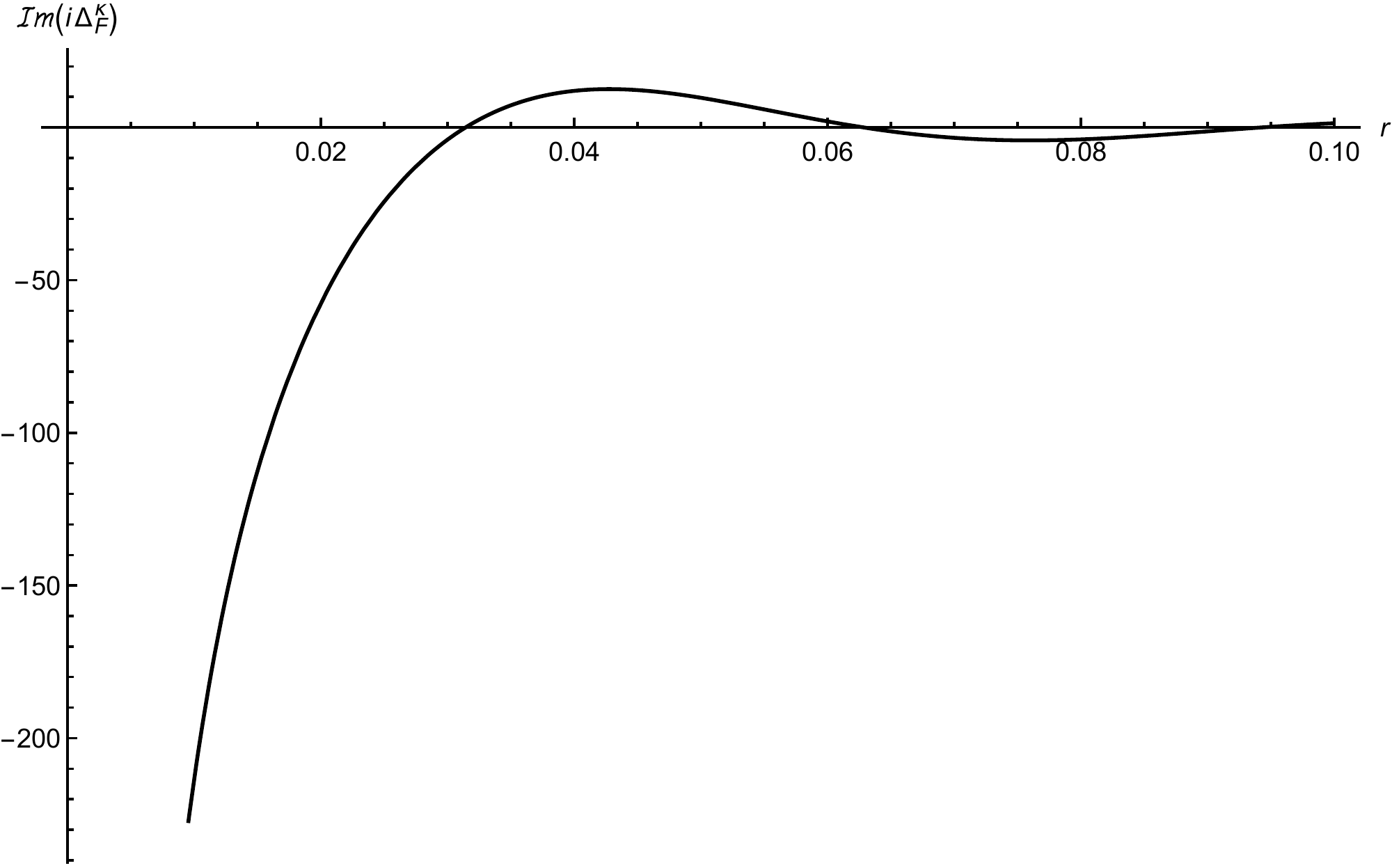}
\caption{{{\footnotesize  Plots of the real (left) and imaginary (right) part of the $ \kappa $-deformed Feynman propagator (\ref{k prop. spacelike}) for spacelike separation $ x^{2}=r^{2}>0 $. In the undeformed case there is no imaginary part. The spacelike separation $ r $ varies on the horizontal axis. In this simulation we have set $ \kappa=10^2 $ and $ m=1 $, so that the ratio is $ m/\kappa=10^{-2} $. We are looking at distances of $ 10 $ Planck lengths $ \kappa^{-1} $ from the light cone, or, equivalently, at distances of $ 10^{-1} $ Compton wavelengths $ m^{-1} $ from the light cone. In the plot of the real part the difference between the standard and the deformed graph is not visually appreciable, being of the order of $ (m/\kappa)^{2} $, and the two graphs overlap.} }}\label{graph.spacelike k prop}
\end{figure}

Given that the term $ i\Pi_{F}^{\kappa}(r) $ is a pure imaginary quantity, the only contribution to real part of the $ \kappa $-deformed Feynman propagator comes from $ i\Delta_{F}(r) $. Such term is real and can be expressed \cite{Greiner} in terms of the Hankel function of the second kind as
\begin{equation}\label{real part k spacelike}
\Re[i\Delta^{\kappa}_{F}(r)]=\frac{i\Delta_{F}(r)}{\sqrt{1+m^{2}/\kappa^{2}}}=\frac{-m}{8\pi\sqrt{1+m^{2}/\kappa^{2}}}\frac{1}{r}H^{(2)}_{1}(-imr) \ .
\end{equation}
Introducing spherical coordinates and performing the angular integration in momentum space, the imaginary part of the $ \kappa $-deformed Feynman propagator (\ref{k prop. spacelike}) reads
\begin{equation}
\begin{split} \Im[i\Delta^{\kappa}_{F}(r)]=i\Pi_{F}^{\kappa}(r)=\frac{- 4i}{\sqrt{1+m^{2}/\kappa^{2}}}\bigg[\int_{0}^{\kappa}\frac{dp}{(2\pi)^{3}}\frac{p}{\omega_{p}}\sinh^{-1}\bigg(\frac{\omega_{p}}{\sqrt{\kappa^{2}-p^{2}}}\bigg)\frac{\sin(p r)}{ r}+& \\ \int_{\kappa}^{\infty}\frac{dp}{(2\pi)^{3}}\frac{p}{\omega_{p}}\cosh^{-1}\bigg(\frac{\omega_{p}}{\sqrt{p^{2}-\kappa^{2}}}\bigg)\frac{\sin(p r)}{ r}\bigg]& \ , \end{split}
\end{equation}
where $ \vert\textbf{p}\vert=p $. In contrast with the standard case, the $ \kappa $-deformed Feynman propagator now possesses both a real and an imaginary part given by the standard Feynman propagator $ i\Delta_{F}(r) $ and by $ i\Pi_{F}^{\kappa}(r) $ respectively. We proceeded to a numerical evaluation of $ i\Pi_{F}^{\kappa}(r) $, and the real and the imaginary part of $ i\Delta^{\kappa}_{F}(r) $ are shown in Figure \ref{graph.spacelike k prop}. 

As in the undeformed case the real part of $ i\Delta^{\kappa}_{F}(r) $ rapidly falls to zero with a scale set by the Compton wavelenght $ m^{-1} $. Indeed, the real part of $ i\Delta^{\kappa}_{F}(r) $ is only modified, w.r.t. the standard case, by a constant multiplicative factor $ (1+m^{2}/\kappa^{2})^{-1/2} $. The leading correction coming from this term is of  order\footnote{Here we have expanded the square root $  (1+\epsilon)^{-1/2} $ w.r.t. the small parameter $ \epsilon=m^{2}/\kappa^{2} \ll 1$, i.e. $(1+\epsilon)^{-1/2}\sim 1-\frac{1}{2}\epsilon+O\big((\epsilon)^{2}\big) \ . $ }  $ (m/\kappa)^{2} $ which does not lead to visually appreciable changes in the first plot in Fig. \ref{graph.spacelike k prop}. The imaginary part of $ i\Delta^{\kappa}_{F}(r) $, absent in the undeformed case, is divergent on the light cone (in accordance with the analysis made in the previous paragraph) and falls to zero oscillating after few Planck lengths $ \kappa^{-1} $ from the light cone.

The fact that the Feynman propagator is non-zero for spacelike distances is a well known quantum mechanical tunneling phenomenon caused by the difficulty to sharply localize a particle in spacetime. However, now, in addition to the localization limit given by the particle's Compton wavelength, we also have an effect generated by the fuzzy nature of spacetime, which results in the non-vanishing additional imaginary contribution of Figure \ref{graph.spacelike k prop} (right).

\subsubsection{Timelike separation}

When the event $x$ is at timelike separation ($ x^{2}<0 $) from the origin we can set $ x^{\mu}=(x^{0}, \textbf{0}) $, so that 
\begin{equation}\label{k prop. timelike}
i\Delta^{\kappa}_{F}(x_{0})=\frac{i\Delta_{F}(x_{0})}{\sqrt{1+m^{2}/\kappa^{2}}}+i\Pi_{F}^{\kappa}(x_{0}) \ .
\end{equation}
The real and the imaginary contributions to $i\Delta^{\kappa}_{F}(x_{0})$ can be identified as follows
\begin{equation}
\begin{split}&\Re\big[i\Delta^{\kappa}_{F}(x_{0})\big]=\Re\bigg[\frac{im}{8\pi\sqrt{1+m^{2}/\kappa^{2}}}\frac{1}{x_{0}}H^{(2)}_{1}(mx_{0})\bigg] \ , \\
&\Im\big[i\Delta^{\kappa}_{F}(x_{0})\big]=\Im\bigg[\frac{im}{8\pi\sqrt{1+m^{2}/\kappa^{2}}}\frac{1}{x_{0}}H^{(2)}_{1}(mx_{0})\bigg]+i\Pi_{F}^{\kappa}(x_{0}) \ , \end{split}
\end{equation}
where we have used the fact that the undeformed Feynman propagator can be expressed in terms of Hankel functions of the second kind and that $ i\Pi_{F}^{\kappa}(x_{0}) $ is a pure imaginary term.

\begin{figure}[h!]
\includegraphics[width=0.45 \textwidth] {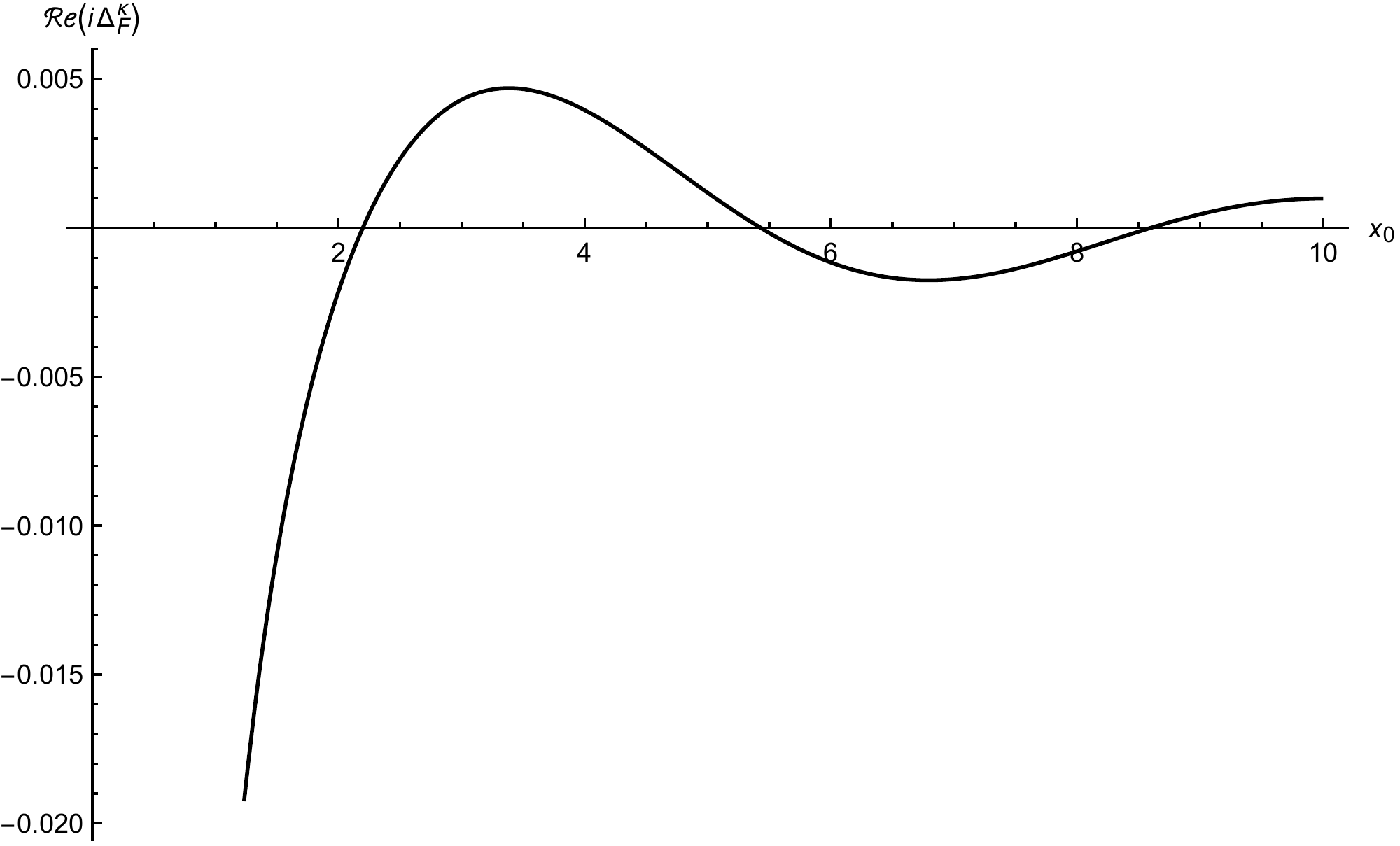}
\includegraphics[width=0.45 \textwidth] {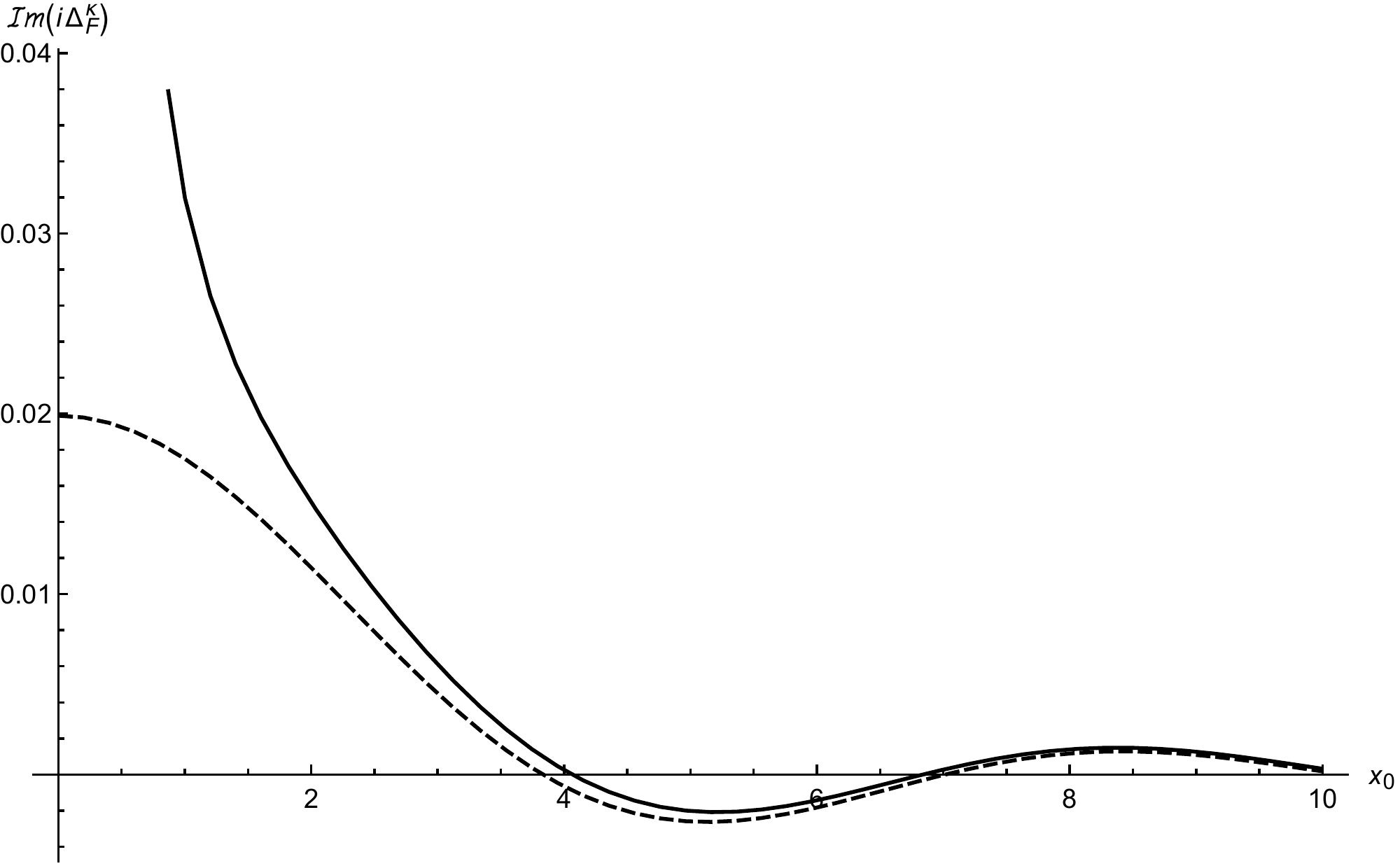}
\caption{{{\footnotesize  Plots of the real (left) and imaginary (right) part of the $ \kappa $-deformed Feynman propagator (\ref{k prop. timelike}) for timelike separation. The timelike interval $ x_{0} $ varies on the horizontal axis. In this simulation we have set $ \kappa=10^2 $ and $ m=1 $, so that the ratio is $ m/\kappa=10^{-2} $. Here we are looking at distances of $ 10^3 $ Planck lengths $ \kappa^{-1} $ from the light cone, or, equivalently, at distances of $ 10 $ Compton wavelengths $ m^{-1} $ from the light cone. In the figure on the right are plotted both the imaginary part of $ i\Delta^{\kappa}_{F}(x_{0}) $ (solid line) and the ordinary undeformed imaginary part of $ i\Delta_{F}(x_{0}) $ (dashed line), while, in the plot of the real part on the left, the difference between the standard and the deformed graph is not visually appreciable, being of the order of $ (m/\kappa)^{2} $, and the two graphs overlap.} }}\label{graph.timelike k prop}
\end{figure}

Introducing spherical polar coordinates and performing the angular integration in momentum space, the term $ i\Pi_{F}^{\kappa}(x_{0}) $ takes the form
\begin{equation}
i\Pi_{F}^{\kappa}(x_{0})=2\bigg[\int_{0}^{\kappa}\frac{dp}{(2\pi)^{3}} \ p^{2} \ I^{A}+\int_{\kappa}^{\infty}\frac{dp}{(2\pi)^{3}} \ p^{2} \ \big(I^{B}_{1}+I_{2}^{B}\big)\bigg] \ ,
\end{equation}
where $ \vert\textbf{p}\vert=p $ and the integrals $ I^{A}, \ I^{B}_{1}$ and $\ I_{2}^{B} $ are defined in (\ref{IA IB1 IB2}). This expression cannot be analytically computed, therefore we proceeded to a numerical evaluation. The resulting real and the imaginary part of $i\Delta^{\kappa}_{F}(x_{0})$ are shown in Figure \ref{graph.timelike k prop}

In Figure \ref{graph.timelike k prop} (right) are shown, in dashed line, the undeformed imaginary part of the propagator and, in solid line, the $ \kappa $-deformed one. The effect of the $ \kappa $-deformation is dominant near the light cone, where the imaginary part of $ i\Delta^{\kappa}_{F}(x_{0}) $ diverges in accordance with the analysis made in the previous paragraphs. For large timelike distances the $ \kappa $-deformed Feynman propagator approaches the standard one, so that it decreases in amplitude while oscillating according to the asymptotic behaviour
\begin{equation}
i\Delta_{F}(x_{0})\sim \mathrm{const}(x_{0}^{2})^{-\frac{3}{4}}e^{-im\vert x_{0}\vert} \ , \ \ \ \ \ \vert x_{0}\vert\rightarrow\infty \ .
\end{equation}

The scale of the oscillations, inside both the future and past lightcone, is set by the Compton wavelengths $ m^{-1} $.
\\

Summarizing, $ \kappa $-deformations modify the real part of the Feynman propagator only by a constant multiplicative factor $ (1+m^{2}/\kappa^{2})^{-1/2} $. Conversely, the imaginary part of the $ \kappa $-deformed Feynman propagator is $ (1+m^{2}/\kappa^{2})^{-1/2} $ times the undeformed one plus an additional contribution $ i\Pi_{F}^{\kappa} $. This contribution is divergent on the light cone, as can be seen from the Figures \ref{graph.spacelike k prop} and \ref{graph.timelike k prop}, for spacelike distances falls to zero oscillating after few Planck length from the light cone and for timelike distances modifies the undeformed propagator as in Figure \ref{graph.timelike k prop} (right).\\ 

Before concluding this Section we would like to note that, in spite of the prominent role that the Feynman propagator plays in the standard local quantum field theory, very few attempts addressing the generalization of the Feynman propagator to non-commutative field theory on $ \kappa $-Minkowski have appeared in literature. Most of the earlier attempts to the study of the $\kappa$-deformed propagator remained just at an exploratory level due to the poorly understood structure of $\kappa$-deformed momentum space at the time. This is the case, for example, of \cite{Kosinski:1999ix} and \cite{Kosinski:2003xx} where the authors introduced a $\kappa$-deformed Feynman propagator and a Pauli-Jordan function by analogy with the undeformed case, simply by replacing the standard relativistic energy-momentum dispersion relation with the $ \kappa $-deformed Casimir similar to \eqref{kCasimir}. More recently a study of the Pauli-Jordan function taking into account the non-trivial de Sitter geometry of $\kappa$-deformed momentum space has appeared in \cite{Mercati:2018hlc}. The approach taken by the authors is to define a field operator and an algebra of creation and annihilation operators, so that the commutator of the fields can be used to derive the Pauli-Jordan via its vacuum expectation value. The main drawback of the canonical approach adopted in \cite{Mercati:2018hlc} is that it relies on the definition of an algebra of creation and annihilation operator which does not take into account the know difficulties in dealing with multiparticle states in a $\kappa$-deformed context (see e.g. \cite{Arzano:2007ef,Arzano:2008bt}). Moreover, using the field operator defined in \cite{Mercati:2018hlc} to construct a $\kappa$-deformed Feynman propagator, one would have that only the simple poles, coming from the delta function of the Casimir, will contribute to the propagator. Our analysis instead shows that the Feynman propagator, in order to be a Green's function of the $ \kappa $-Klein-Gordon equation, must contain also the contributions from the branch cuts.

\section{From non-commutative to non-local fields}

In the previous Section we have seen how the non-trivial features of $ \kappa $-deformed propagation (\ref{propagation law 2}) can be understood in terms of an effective spacetime fuzziness in the UV. We also pointed out how the deformed field propagation can be equivalently viewed as a propagation on classical Minkowski spacetime, i.e. with no limitations in localizing a source, mediated by the $ \kappa $-deformed Feynman propagator (\ref{kappa deformed feynman propagator}).

In this Section we show how such interpretation of  $ \kappa $-deformed propagation, based on a classical Minkowski spacetime, can be formulated in terms of a non-local field theory.  We will also show how the $ \kappa $-deformed Feynman propagator can be related to the vacuum expectation values of products of this non-local field operator. We will find that the standard relation between the Feynman propagator and the time-ordered two-point function holds only in the sub-Planckian sector, while, for trans-Planckian momenta, the $ \kappa $-deformed Feynman propagator will be shown to have the form of the Hadamard (anti-commutator) two-point function. We will comment on a possible physical interpretation of such feature at the end of this Section.

As observed in \cite{Freidel:2006gc}, \cite{KowalskiGlikman:2009zu}, the free $ \kappa $-Minkowski scalar field theory can be recast in the form of a non-local scalar field theory on ordinary Minkowski space. In these works it was shown how the non-local character of the $ \kappa $-Minkowski theories, already noticed in \cite{Kosinski:1999ix} for the bicrossproduct basis, is present also for the classical basis and it is encoded in a non-local star product. Let us notice that the non-local character of the $ \kappa $-Minkowski free scalar action could have already been noted by considering its expression in momentum space (\ref{action in momentum space}) and the explicit form of the Fourier transform \eqref{explicit fourier transform}, so that \eqref{action in momentum space} can be written as
\begin{equation}
S_{free}=\int d^{4}x   d^{4}x' \phi^{\ast}(x)\big(-\square+m^{2}\big)\phi(x')V(x,x') \,, 
\end{equation}
where the non-local term $V(x,x') $ is given by
\begin{equation}
V(x,x')=\int \frac{d^{4}p \ \theta(\kappa^{2}-p^{2})}{(2\pi)^{4}}\sqrt{1-p^{2}/\kappa^{2}} \ e^{-ip(x'-x)} \ .
\end{equation}
Alternatively, making use of the classical basis Weyl map $ \Omega_{c} $ and its associated star product $ \star $ (see Section \ref{wmsp}), the $ \kappa $-Poincar\'e invariant action (\ref{action k scalar field}) can be expressed as
\begin{equation}
S_{free}=\int d^{4}x \ \big[(\partial_{\mu}\phi)^{\dagger} \star (\partial^{\mu}\phi)+m^{2}\phi^{\dagger}\star \phi\big]=\int d^{4}x \ \big[(\partial_{\mu}\phi)^{\ast} \sqrt{1+\square/\kappa^{2}} \  (\partial^{\mu}\phi)+m^{2}\phi^{\ast}\sqrt{1+\square/\kappa^{2}} \ \phi\big] \ .
\end{equation}
This action is manifestly invariant under $ \kappa $-Poincar\'e trasformations which in the classical basis, as mentioned above, are just the standard ones. Taking the formal series expansion in powers of the d'Alembertian for the star product term, the action has infinitely many derivatives 
\begin{equation}\label{action series}
S_{free}=\int d^{4}x \sum_{n=0}^{\infty} a_{n} \big[ \partial_{\mu}\phi^{\ast} \square^{n}\partial^{\mu}\phi +m^{2}\phi^{\ast}\square^{n}\phi\big] \ ,
\end{equation}
where $ a_{n}\propto \kappa^{-2n} $. Varying the action (\ref{action series}) with respect to the field and its complex conjugate we get
\begin{equation}\label{series expansion equation of motion}
\sum_{n=0}^{\infty} a_{n}\big[-\square^{n+1}+m^{2}\square^{n}\big]\phi=0 \  ,
\end{equation}
and similary for $ \phi^{\ast} $, from which, summing up the series, one obtains the non-local equation of motion 
\begin{equation}\label{eq mink.}
\sqrt{1+\square/\kappa^{2}}\big(\square-m^{2}\big) \phi=0 \,  , %\ \ \ \ \ \ \ \ \  \sqrt{1+\square/\kappa^{2}}\big(\square-m^{2}\big) \phi^{\ast}=0 \  .
\end{equation}
and an identical one for the complex conjugate field $\phi^{\ast}$. These equations of motion involve non-local pseudo-differential operators, specifically, they contain fractional powers of the d'Alembertian. In order to solve them, we will make use of the methods developed in \cite{doAmaral:1992td, Barci:1995ad,Barci:1996br} and employed, for example, in \cite{Belenchia:2014fda} to study the non-local effective field theory emerging in a ``mesoscopic" regime of casual sets.  The general form of a Lorentz invariant non-local pseudo-differential equation is
\begin{equation}\label{pseudo diff eq.}
f\big(\square\big)\phi=0 \ ,
\end{equation}
where, as for the equation of motion (\ref{series expansion equation of motion}), the function $ f\big(\square\big) $ cannot be expanded in a finite series. A general solution of (\ref{pseudo diff eq.}) can be written as 
\begin{equation}\label{k solution 1}
\phi(x)=\dfrac{1}{2\pi i}\int \frac{d^{3}p}{(2\pi)^{3/2}} \bigintssss_{\sum_{i}\Gamma_{i}}dp_{0} \ e^{ipx} \ \big[\frac{1}{f(-p^{2})}\big] \ a(p_{0},\textbf{p}) \ ,
\end{equation}
where $ a(p_{0},\textbf{p}) $ is an entire analytic function, i.e. a complex-valued function that is holomorphic at all points over the whole complex plane, and the $\Gamma_{i}  $'s are paths that encircle, in the complex $ p_{0} $ plane, all the singularities of $ 1/f(-p^{2}) $. Indeed, rewriting the solution (\ref{k solution 1}) as 
\begin{equation}\label{field solution k}
\phi(x)=\int \frac{d^{3}p}{(2\pi)^{3/2}} \ e^{i\textbf{p}\cdot\textbf{x}} \ \phi_{\textbf{p}}(x_{0}) \ ,
\end{equation}
where
\begin{equation}\label{field solution p0 part}
\phi_{\textbf{p}}(x_{0})=\dfrac{1}{2\pi i}\bigintssss_{\sum_{i}\Gamma_{i}}dp_{0} \ e^{-ip_{0}x_{0}} \ \big[\frac{1}{f(-p^{2})}\big] \ a(p_{0},\textbf{p}) \ ,
\end{equation}
the condition (\ref{pseudo diff eq.}) becomes
\begin{equation}\label{eq for p0 part}
f\big(-\partial_{0}^{2}-\textbf{p}^{2}\big)\phi_{\textbf{p}}(x_{0})=0 \ ,
\end{equation}
and when we apply the operator $ f\big(-\partial_{0}^{2}-\textbf{p}^{2}\big) $ on (\ref{field solution p0 part}) a factor $ f(-p^{2}) $ that cancels all the poles and cuts is produced inside the integral, so that the paths $ \Gamma_{i} $ can now be deformed to a point giving vanishing contributions. In our case the function $ f(-p^{2}) $ can be read off the $ \kappa $-deformed equations of motion (\ref{eq mink.}) and is given by
\begin{equation}\label{non-local function f}
f_{\kappa}(-p^{2})=\sqrt{1-p^{2}/\kappa^2} \big(-p^{2}-m^{2}\big)\ .
\end{equation}
An important point to notice is that the $ \kappa $-deformed contribution $ \sqrt{1-p^{2}/\kappa^2} $ to the function $ f_{\kappa}(-p^{2}) $ is just the term $ \vert p_{4}\vert/\kappa $ appearing at the denominator of the $ \kappa $-Poincar\'e momentum space integration measure $ d\bar{\mu}(p)=d^{4}p\theta(\kappa^{2}-p^{2})\kappa/\vert p_{4}\vert $. Accordingly, the isolated singularities and branch cuts associated with the function $ 1/f_{\kappa}(-p^{2}) $ are just the ones we have already seen in Section 3.3 when solving the $ p_{0} $ integral of the $ \kappa $-deformed Feynman propagator. We will thus split the analysis of the solutions associated to (\ref{eq mink.}) in two parts depending on whether one considers the sub-Planckian or the trans-Planckian sector. The position of the branch cuts in the complex $ p_{0} $ plane are different in these two case; a cut on the immaginary $ p_{0} $ axis when $ \vert\textbf{p}\vert<\kappa $ and a cut on the real $ p_{0} $ axis when $ \vert\textbf{p}\vert>\kappa $ (see Fig. (\ref{kpath p0 1}) and (\ref{kpath p0 4})). The solution (\ref{field solution k}) can thus be rewritten as
\begin{equation}\label{field A+B}
\phi(x)=\phi^{A}(x)+\phi^{B}(x)=\int_{\vert\textbf{p}\vert<\kappa} \frac{d^{3}p}{(2\pi)^{3/2}} \ e^{i\textbf{p}\cdot\textbf{x}} \ \phi^{A}_{\textbf{p}}(x_{0})+\int_{\vert\textbf{p}\vert>\kappa} \frac{d^{3}p}{(2\pi)^{3/2}} \ e^{i\textbf{p}\cdot\textbf{x}} \ \phi^{B}_{\textbf{p}}(x_{0})  \ ,
\end{equation}
and similary for $ \phi^{\ast}(x) $. Our goal in the next Section will be to find a relationship between the $ \kappa $-deformed Feynman propagator found in Section 3 and vacuum expectation values of the field operator \eqref{field A+B} above. Let us recall that, in analogy with \eqref{field A+B}, the propagator (\ref{kappa deformed feynman expand 2}) exhibits a similar splitting in sub-plankian and trans-planckian contributions 
\begin{equation}\label{Feynman propagator split}
i\Delta_{F}^{\kappa}(x)=i\Delta_{F}^{\kappa}(x)\big\vert_{A}+i\Delta_{F}^{\kappa}(x)\big\vert_{B}=i\int_{\vert\textbf{p}\vert<\kappa} \frac{d^{3}p}{(2\pi)^{4}}e^{i\textbf{p}\cdot\textbf{x}} \ \mathcal{I}^{A}_{\textbf{p}}(x_{0})+i\int_{\vert\textbf{p}\vert>\kappa} \frac{d^{3}p}{(2\pi)^{4}}e^{i\textbf{p}\cdot\textbf{x}} \ \mathcal{I}^{B}_{\textbf{p}}(x_{0}) \ ,
\end{equation}
due to the restriction imposed by the condition $\kappa^{2}-p^{2}>0 $ on momentum space. In particular, we should notice that, since in the sub-Planckian region $ A $ the $ p_{0} $ range of integration in the propagator remains unaffected (see (\ref{region a b})), we expect a standard relation between $ i\Delta_{F}^{\kappa}(x)\big\vert_{A} $ and the time-ordered two-point function of $ \phi^{A} $. The same, however, can not be said for the trans-Planckian sector $ B $. Here the condition $ \kappa^{2}-p^{2}>0 $ deforms the $ p_{0} $ range of integration ($\vert p_{0}\vert> \Omega_{B}$), so we should expect a different combination of two-point functions of $ \phi^{B} $ to be related to $ i\Delta_{F}^{\kappa}(x)\big\vert_{B} $. In what follows we will study this issue in detail.

\section{The $\kappa$-deformed propagator as a non-local two-point function}\label{Sec On the connection}

In this Section we characterize the $ \kappa $-deformed Feynman propagator in terms of the vacuum expectation values of the non-local complex scalar field on ordinary Minkowski spacetime introduced in the previous Section. In order to do so we proceed to quantize the non-local field \eqref{field solution k} using the techniques first developed in \cite{doAmaral:1992td}. Once again we focus separately on the sub-Planckian and trans-Planckian cases.

\subsection{Sub-Planckian momenta: time-ordered two-point function}

\begin{figure}[h!]
\centering
\includegraphics[width=0.55 \textwidth] {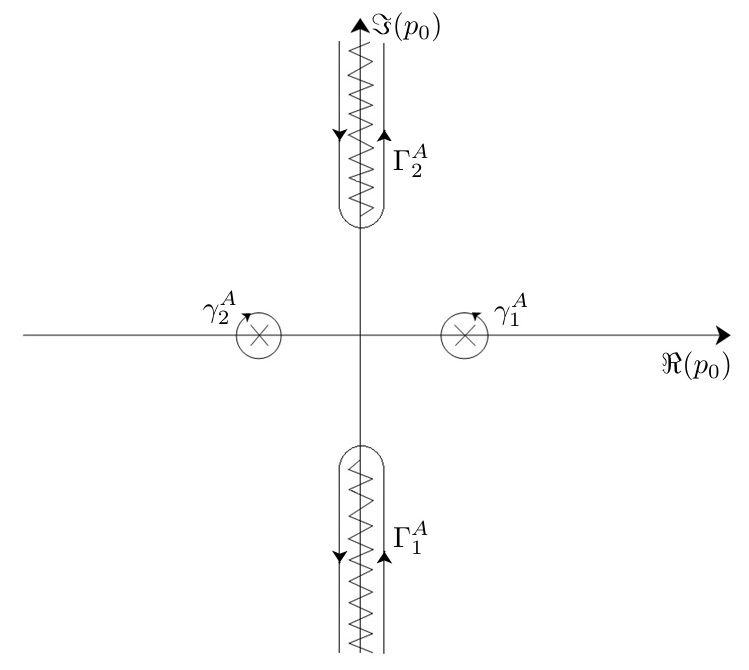}
\caption{{ {\footnotesize Singularity structure in the region $ A $. There are two simple poles at $ p_{0}=\pm\omega_{\textbf{p}} $ and two branch cuts on the imaginary axis, one from $ i\Omega_{A} $ to $ i\infty $ and one from $ -i\Omega_{A} $ to $ -i\infty $.}}} \label{tagli A}
\end{figure}

In the region of sub-Planckian momenta the function $ f_{\kappa} $ of the equation (\ref{non-local function f}) takes the form
\begin{equation}\label{f region A}
f_{\kappa}(p_{0},\textbf{p})=\frac{2\pi i}{\kappa}\sqrt{p_{0}^{2}+\Omega_{A}^{2}}\big(p_{0}^{2}-\omega_{\textbf{p}}^{2}\big) \ ,
\end{equation}
where, for future convenience, we have incorporated also the term $ 2\pi i $ in the definition of $ f_{\kappa} $ and, as in Section 3, $ \Omega_{A}=\sqrt{\kappa^{2}-\textbf{p}^{2}} $ and $ \omega_{\textbf{p}}=\sqrt{\textbf{p}^{2}+m^{2}} $. Considering equations (\ref{field solution p0 part}), (\ref{field A+B}) and (\ref{f region A}), it is then possible to express the restriction of the field $ \phi $ to the region $ A $, taking into account the singularity structure of $ 1/f_{\kappa}$, as
\begin{equation}
\phi^{A}(x)=\int_{\vert\textbf{p}\vert<\kappa} \frac{d^{3}p }{(2\pi)^{3/2}} e^{i\textbf{p}\cdot\textbf{x}} \ \phi^{A}_{\textbf{p}}(x_{0}) \ ,
\end{equation}
with 
\begin{equation}\label{expansion field A}
\phi^{A}_{\textbf{p}}(x_{0})=\dfrac{1}{2\pi i}\sum_{i=1}^{2}\int_{\Gamma^{A}_{i}+\gamma^{A}_{i}}\frac{dp_{0} \ \kappa}{\sqrt{p_{0}^{2}+\Omega_{A}^{2}}} \ \frac{e^{-ip_{0}x_{0}}}{p_{0}^{2}-\omega_{\textbf{p}}^{2}} \ a(p_{0},\textbf{p}) \ ,
\end{equation}
where the integration contours $ \Gamma^{A}_{i}$ and $\gamma^{A}_{i} $ are shown in Figure \ref{tagli A}. We can write (\ref{expansion field A}) explicitly as 
\begin{multline}\label{espansione field A}
\phi^{A}_{\textbf{p}}(x_{0})=\int_{-i\infty}^{-i\Omega_{A}}dp_{0} \ \Delta_{\Gamma_{A}}\bigg[\frac{1}{f_{\kappa}(p_{0},\textbf{p})}\bigg] \ e^{-ip_{0}x_{0}} \ a(p_{0},\textbf{p}) \ +\int_{-i\infty}^{-i\Omega_{A}}dp_{0} \ \Delta_{\Gamma_{A}}\bigg[\frac{1}{f_{\kappa}(p_{0},\textbf{p})}\bigg] \ e^{ip_{0}x_{0}} \ a(-p_{0},\textbf{p}) \ + \\
+\frac{1}{\sqrt{1+m^{2}/\kappa^{2}}}\dfrac{e^{-i\omega_{\textbf{p}}x_{0}}}{2\omega_{\textbf{p}}} \ a(\omega_{\textbf{p}},\textbf{p}) \ + \ \frac{1}{\sqrt{1+m^{2}/\kappa^{2}}}\dfrac{e^{i\omega_{\textbf{p}}x_{0}}}{2\omega_{\textbf{p}}} \ a(-\omega_{\textbf{p}},\textbf{p}) \ , \ \ \ \ \ \ \ \ \ \ \ \ \ \ \ \ \
\end{multline} 
where the functional $ \Delta_{\Gamma_{A}}\big[1/f_{\kappa}(p_{0},\textbf{p})\big] $ is the discontinuity functional at the branch cut of $ 1/f_{\kappa} $, which in our case is just
\begin{equation}\label{discontinuity cut A}
\Delta_{\Gamma_{A}}\bigg[\frac{1}{f_{\kappa}(p_{0},\textbf{p})}\bigg]=-\frac{2}{2\pi i}\frac{\kappa}{\sqrt{p_{0}^{2}+\Omega_{A}^{2}}} \frac{1}{p_{0}^{2}-\omega_{\textbf{p}}^{2}} \, .
\end{equation}

The quantum counterpart of the field (\ref{espansione field A}) is obtained by promoting the entire analityc coefficients $ a(p) $ and $ a(-p) $ to annihilation  and creation operators respectively. Since we are dealing with a complex field, the function $ a(p) $ will be promoted to the annihilation operator for particles while $ a(-p) $ to the creation operator $ b^{\dagger}(p) $ of antiparticles. Following \cite{doAmaral:1992td} we note that, in order to have a consistent quantization scheme, the creation and annihilation operators must satisfy the commutation rules 
\[
\Delta\bigg[\frac{1}{f_{\kappa}(-p^{2})}\bigg] \big[a(p), a^{\dagger}(q)\big]=\delta(p-q) \ ,
\]
\begin{equation}\label{commutation relation 1}
\Delta\bigg[\frac{1}{f_{\kappa}(-p^{2})}\bigg] \big[b(p), b^{\dagger}(q)\big]=\delta(p-q) \ ,
\end{equation}
with all the other commutators being zero. The discontinuity functional $ \Delta\big[1/f_{\kappa}(-p^{2})\big] $ involved in the commutation relations (\ref{commutation relation 1}) obviously depends on which singularity one considers; for instance, the discontinuity functional associated to the branch cut is the one appearing in (\ref{discontinuity cut A}), while for the simple pole one has
\begin{equation}
\Delta_{\gamma_{A}}\bigg[\frac{1}{f_{\kappa}(-p^{2})}\bigg]=\lim_{\varepsilon\rightarrow 0}\bigg[\bigg(\dfrac{1}{f_{\kappa}((p_{0}+i\varepsilon)^{2}-\textbf{p}^{2})}-\dfrac{1}{f_{\kappa}((p_{0}-i\varepsilon)^{2}-\textbf{p}^{2})}\bigg)\bigg\vert_{\gamma_{A}}\bigg] \ ,
\end{equation}
and with simple manipulations
\begin{equation}
\Delta_{\gamma_{A}}\bigg[\frac{1}{f_{\kappa}(-p^{2})}\bigg]=\frac{1}{2\pi i}\frac{2\pi i}{\sqrt{1-p^{2}/\kappa^{2}}}\lim_{\varepsilon\rightarrow 0}\bigg(\frac{1}{\pi}\dfrac{\varepsilon}{\varepsilon^{2}+(-p^{2}-m^{2})^{2}}\bigg)=\frac{\delta(p^{2}+m^{2})}{\sqrt{1+m^{2}/\kappa^{2}}} \ .
\end{equation}
The algebra of the annihilation and creation operators for the field excitations with four-momentum associated to the simple pole are then simply $ \sqrt{1+m^{2}/\kappa^{2}} $ times the usual ones
\[
\big[a(\omega_{\textbf{p}},\textbf{p}),a^{\dagger}(\omega_{\textbf{q}},\textbf{q})\big]=2\omega_{\textbf{p}}\sqrt{1+m^{2}/\kappa^{2}} \  \delta(\textbf{p}-\textbf{q}) \ , 
\]
\begin{equation}\label{commutatore A polo}
\big[b(\omega_{\textbf{p}},\textbf{p}),b^{\dagger}(\omega_{\textbf{q}},\textbf{q})\big]=2\omega_{\textbf{p}} \sqrt{1+m^{2}/\kappa^{2}} \ \delta(\textbf{p}-\textbf{q}) \ ,
\end{equation}
while for excitations with four-momentum belonging to the cut one obtains
\[
\big[a(p), a^{\dagger}(q)\big]\bigg\vert_{p,q \in \Gamma_{A}}= \frac{\delta(p-q)}{\Delta_{\Gamma_{A}}\big[1/f(-p^{2})\big]} \ ,
\]
\begin{equation}\label{commmutatore A taglio 2}
\big[b(p), b^{\dagger}(q)\big]\bigg\vert_{p,q \in \Gamma_{A}}= \frac{\delta(p-q)}{\Delta_{\Gamma_{A}}\big[1/f(-p^{2})\big]} \ .
\end{equation}

We now have all the tools needed to compute the time-ordered two-point functions in the region A as
\begin{equation}
\begin{split}\langle 0\vert T\big\lbrace\phi^{A}(x)(\phi^{ A}(y))^{\ast}\big\rbrace\vert 0\rangle=&\theta(x_{0}-y_{0})\int_{\vert\textbf{p}\vert,\vert\textbf{q}\vert<\kappa} \frac{d^{3}p \ d^{3}q}{(2\pi)^{3}} \ e^{i\textbf{p}\cdot\textbf{x}} e^{-i\textbf{q}\cdot\textbf{y}}\ \langle 0\vert \phi^{A}_{\textbf{p}}(x_{0})(\phi^{A}_{\textbf{q}}(y_{0}))^{\ast}\vert 0\rangle \ +  \\
+ &\theta(y_{0}-x_{0})\int_{\vert\textbf{p}\vert,\vert\textbf{q}\vert<\kappa} \frac{d^{3}p \ d^{3}q}{(2\pi)^{3}} \ e^{-i\textbf{p}\cdot\textbf{x}}e^{i\textbf{q}\cdot\textbf{y}} \ \langle 0\vert (\phi^{A}_{\textbf{q}}(y_{0}))^{\ast}\phi^{A}_{\textbf{p}}(x_{0})\vert 0\rangle \ , \end{split}
\end{equation}
and, making use of the equations (\ref{espansione field A}), (\ref{commutatore A polo}) and (\ref{commmutatore A taglio 2}), we get\footnote{In deriving the integral inside the square brackets we have made the changes of variable $ p_{0}\rightarrow\pm iz $.}
\begin{multline}\label{t-ordered A}
\langle 0\vert T\big\lbrace\phi^{A}(x)(\phi^{A}(y))^{\ast}\big\rbrace\vert 0\rangle=\int_{\vert\textbf{p}\vert<\kappa} \frac{d^{3}p}{(2\pi)^{3}} \ e^{i\textbf{p}\cdot(\textbf{x}-\textbf{y})} \  \bigg[\dfrac{e^{-i\omega_{\textbf{p}}\vert x_{0}-y_{0}\vert}}{2\omega_{\textbf{p}}\sqrt{1+m^{2}/\kappa^{2}}} \ +\frac{2}{2\pi i}\int_{\Omega_{A}}^{\infty}\dfrac{dz \ \kappa}{\sqrt{z^{2}-\Omega_{A}^{2}}}\dfrac{e^{-z\vert x_{0}-y_{0}\vert}}{z^{2}+\omega_{\textbf{p}}^{2}} \ \bigg].
\end{multline}
The term in square brackets is just $ i/2\pi $ times the integral $ \mathcal{I}^{A}_{\textbf{p}} $ defined in (\ref{risultato A p0}). Accordingly, this expression is equal to the restriction $ i\Delta^{\kappa}_{F}(x-y)\big\vert_{A} $ of the $ \kappa $-deformed Feynman propagator to the region of sub-Planckian momenta, as can be easily seen confronting the relations (\ref{Feynman propagator split}) and (\ref{t-ordered A})
\begin{equation}\label{t-ordered A con I}
\langle 0\vert T\big\lbrace\phi^{A}(x)(\phi^{A}(y))^{\ast}\big\rbrace\vert 0\rangle=i\int_{\vert\textbf{p}\vert<\kappa} \dfrac{d^{3}p}{(2\pi)^{4}} e^{i\textbf{p}\cdot(\textbf{x}-\textbf{y})} \ \mathcal{I}_{\textbf{p}}^{A}(x_{0}-y_{0})\equiv i\Delta^{\kappa}_{F}(x-y)\big\vert_{A} \,.
\end{equation}
This shows that for sub-Planckian modes, as aspected, no changes occur in the standard relation between the Feynman propagator and the vacuum expectation values of the T-product of fields.

\subsection{Trans-Planckian momenta: the Hadamard function}

\begin{figure}[h!]
\centering
\includegraphics[width=0.55 \textwidth] {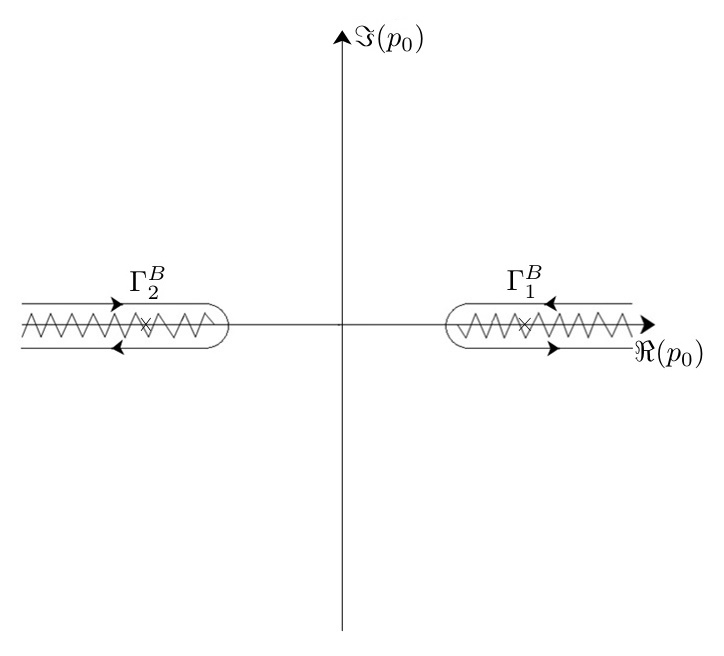}
\caption{{{\footnotesize  Singularity structure in the region $ B $. There are two branch cuts on the real axis, one from $ \Omega_{B} $ to $ +\infty $ and one from $ -\Omega_{B} $ to $ -\infty $. The two simple poles at $ p_{0}=\pm\omega_{\textbf{p}} $ are just over the branch cuts since when $ \vert\textbf{p}\vert>\kappa $ then $ \omega_{\textbf{p}}>\Omega_{B}$.}}} \label{tagli B}
\end{figure}

We now derive the combination of non-local two-point functions which reproduces the restriction $ i\Delta^{\kappa}_{F}(x-y)\big\vert_{B} $ of the $ \kappa $-deformed Feynman propagator to the region of trans-Planckian momenta. Let us start from the $ \kappa $-deformed Feynman propagator
\begin{equation}
i\Delta^{\kappa}_{F}(x-y)=i\int \dfrac{d^{4}p \ \theta(\kappa^{2}-p^{2})}{(2\pi)^{4}\sqrt{1-p^{2}/\kappa^{2}}}\dfrac{e^{ip(x-y)}}{p_{0}^{2}-\omega_{\textbf{p}}^{2}+i\varepsilon} \,,
\end{equation}
which we can rewrite using the integral expression for the step function $\theta(\kappa^{2}-p^{2})$ as
\begin{equation}
i\Delta^{\kappa}_{F}(x-y)=i\int_{-\kappa^{2}}^{+\infty}d\mu^{2}\int \dfrac{d^{4}p \ \delta(p^{2}+\mu^{2})}{(2\pi)^{4}\sqrt{1-p^{2}/\kappa^{2}}} \dfrac{e^{ip(x-y)}}{p_{0}^{2}-\omega_{\textbf{p}}^{2}+i\varepsilon} \,.
\end{equation}
The restriction to spatial momenta bigger than $ \kappa $ is given by
\begin{equation}\label{k spectral B}
i\Delta^{\kappa}_{F}(x-y)\big\vert_{B}=i\int_{-\kappa^{2}}^{+\infty}d\mu^{2}\int_{ \vert\textbf{p}\vert>\kappa } \dfrac{d^{4}p \ \delta(p^{2}+\mu^{2})}{(2\pi)^{4}\sqrt{1-p^{2}/\kappa^{2}}} \dfrac{e^{ip(x-y)}}{p_{0}^{2}-\omega_{\textbf{p}}^{2}+i\varepsilon}  \ .
\end{equation}
For trans-Planckian momenta the delta function $ \delta(p^{2}+\mu^{2}) $ can be expanded with respect to the roots of $ p_{0} $ as\footnote{Indeed it is only for  $ \vert\textbf{p}\vert>\kappa $ that $ \omega_{\textbf{p}}^{\mu}\in\mathbb{R} $ for all $ \mu^{2}\in [-\kappa^{2},+\infty) $.}
\begin{equation}
\delta(p^{2}+\mu^{2})= \dfrac{\delta(p_{0}-\omega_{\textbf{p}}^{\mu})+\delta(p_{0}+\omega_{\textbf{p}}^{\mu})}{2\omega_{\textbf{p}}^{\mu}} \ , \ \ \ \ \ \ \omega_{\textbf{p}}^{\mu}=\sqrt{\textbf{p}^{2}+\mu^{2}} \ ,
\end{equation} 
and the integral over $ p_{0} $ in (\ref{k spectral B}) can be trivially carried out
\begin{equation}\label{kkjo}
i\Delta^{\kappa}_{F}(x-y)\big\vert_{B}=i\int_{-\kappa^{2}}^{+\infty}\frac{d\mu^{2} \ \kappa}{(2\pi)\sqrt{\kappa^{2}+\mu^{2}}}\frac{1}{\mu^{2}+m^{2}+i\varepsilon}\bigg[\int_{\vert\textbf{p}\vert>\kappa }\dfrac{d^{3}p \ e^{i\textbf{p}(\textbf{x}-\textbf{y})}}{(2\pi)^{3} \ 2\omega_{\textbf{p}}^{\mu}}\big(  e^{-i\omega_{\textbf{p}}^{\mu}(x_{0}-y_{0})}+e^{i\omega_{\textbf{p}}^{\mu}(x_{0}-y_{0})} \big)\bigg] \ .
\end{equation}
The term in square brackets is just the restriction to $ \vert\textbf{p}\vert>\kappa $ of the Feynman propagator $ i\Delta_{F}(x-y;\mu^{2}) $ minus the anti-propagator $ i\Delta_{AF}(x-y;\mu^{2})$ or, equivalently, the Hadamard (anti-commutator) function $ G^{(1)}(x-y;\mu^{2}) $ of an ordinary scalar field of mass $ \mu $ 
\begin{equation}
G^{(1)}(x-y;\mu^{2})\big\vert_{B}=i\Delta_{F}(x-y;\mu^{2})\big\vert_{B}-i\Delta_{AF}(x-y;\mu^{2})\big\vert_{B}=\int_{\vert\textbf{p}\vert>\kappa }\dfrac{d^{3}p \ e^{i\textbf{p}(\textbf{x}-\textbf{y})}}{(2\pi)^{3} \ 2\omega_{\textbf{p}}^{\mu}}\big(  e^{-i\omega_{\textbf{p}}^{\mu}\vert x_{0}-y_{0}\vert}+e^{i\omega_{\textbf{p}}^{\mu}\vert x_{0}-y_{0}\vert} \big) \,.
\end{equation}
We can thus write \eqref{kkjo} in terms of a K{\"a}llen-Lehmann-like spectral representation 
\begin{equation}\label{k spectral B 2}
i\Delta^{\kappa}_{F}(x-y)\big\vert_{B}=\int_{-\kappa^{2}}^{+\infty}d\mu^{2} \ \sigma(\mu^{2}) \ \frac{1}{2}\bigg[i\Delta_{F}(x-y;\mu^{2})\big\vert_{B}-i\Delta_{AF}(x-y;\mu^{2})\big\vert_{B}\bigg] \, ,
\end{equation}
where the spectral function $ \sigma(\mu^{2}) $ is the discontinuity at the cut from $ -\kappa^{2} $ to $ +\infty $ of the function $ 1/f_{\kappa}(\mu^{2})=\big(2\pi i(\mu^{2}+m^{2})\sqrt{1+\mu^{2}/\kappa^2}\big)^{-1}  $
\begin{equation}\label{spectral function}
\sigma(\mu^{2})=-\frac{2}{2\pi i}\frac{\kappa}{\sqrt{\kappa^{2}+\mu^{2}}}\frac{1}{\mu^{2}+m^{2}+i\varepsilon} \ .
\end{equation}
The analysis above suggest that the $ \kappa $-deformed Feynman propagator for trans-Planckian momenta should take the form of half the Hadamard function
\begin{equation}\label{combin. wightman}
\frac{1}{2}\langle 0\vert \lbrace\phi^{B}(x),(\phi^{B}(y))^{\ast}\rbrace\vert 0\rangle\equiv\frac{1}{2}\bigg[\langle 0\vert \phi^{B}(x)(\phi^{B}(y))^{\ast}\vert 0\rangle +\langle 0\vert (\phi^{B}(y))^{\ast}\phi^{B}(x)\vert 0\rangle\bigg] \ .
\end{equation}
In order to check this explicitly we evaluate the anti-commutator of the field in region $ B $. For trans-Planckian momenta the function $ f_{\kappa} $ takes the form $ f_{\kappa}(p_{0},\textbf{p})=\frac{2\pi i}{\kappa}\sqrt{p_{0}^{2}-\Omega_{B}^{2}}(p_{0}^{2}-\omega_{\textbf{p}}^{2}) $, where $ \Omega_{B}=\sqrt{\textbf{p}^{2}-\kappa^{2}} $. The singularity structure of $ 1/f_{\kappa} $ is shown in Figure \ref{tagli B}. Taking into account equations (\ref{field solution p0 part}) and (\ref{field A+B}) we can write the field as 
\begin{equation}\label{k field p0 B}
\begin{split}\phi^{B}_{\textbf{p}}(x_{0})=\dfrac{1}{2\pi i}\sum_{i=1}^{2}\int_{\Gamma^{B}_{i}}\frac{dp_{0} \ \kappa}{\sqrt{p_{0}^{2}-\Omega_{B}^{2}}} \ \frac{e^{-ip_{0}x_{0}}}{p_{0}^{2}-\omega_{\textbf{p}}^{2}} \ a(p_{0},\textbf{p})=&\int_{\Omega_{B}}^{\infty}dp_{0} \ \Delta_{\Gamma_{B}}\big[\frac{1}{f_{\kappa}(p_{0},\textbf{p})}\big] \ e^{-ip_{0}x_{0}} \ a(p_{0},\textbf{p}) \ +  \\
+&\int_{\Omega_{B}}^{\infty}dp_{0} \ \Delta_{\Gamma_{B}}\big[\frac{1}{f_{\kappa}(p_{0},\textbf{p})}\big] \ e^{ip_{0}x_{0}} \ a(-p_{0},\textbf{p}) \ , \end{split} 
\end{equation}
where the discontinuity functional at the cut of $ 1/f_{\kappa} $ is 
\begin{equation}\label{discontinuity cut B}
\Delta_{\Gamma_{B}}\big[\frac{1}{f_{\kappa}(p_{0},\textbf{p})}\big]=-\frac{2}{2\pi i}\frac{\kappa}{\sqrt{p_{0}^{2}-\Omega_{B}^{2}}} \frac{1}{p_{0}^{2}-\omega_{\textbf{p}}^{2}} \ .
\end{equation}
Following the procedure already employed in the region $ A $, the quantum counterpart of the field (\ref{k field p0 B}) and of its complex conjugate, are obtained by promoting the entire analytic functions $ a(p) $ and $ a(-p) $ to annihilation and creation operators of particles and antiparticles respectively. The commutation rules for annihilation and creation operators of excitations with four-momentum belonging to the cut are then
\begin{equation}\label{commmutatore B taglio 2}
\big[a(p), a^{\dagger}(q)\big]\bigg\vert_{p,q \in \Gamma_{B}}= \frac{\delta(p-q)}{\Delta_{\Gamma_{B}}\big[1/f(-p^{2})\big]} \ ,  \ \ \ \ \ \big[b(p), b^{\dagger}(q)\big]\bigg\vert_{p,q \in \Gamma_{B}}= \frac{\delta(p-q)}{\Delta_{\Gamma_{B}}\big[1/f(-p^{2})\big]} \ ,  
\end{equation}
with all the others commutators being zero.
Using the equations (\ref{k field p0 B})$-$(\ref{commmutatore B taglio 2}), one can now compute the two-point functions for trans-Planckian momenta as
\begin{equation}\label{wightman functions region B}
\begin{split}\langle 0\vert \phi^{B}(x)(\phi^{B}(y))^{\ast}\vert 0\rangle&=\int_{\vert\textbf{p}\vert,\vert\textbf{q}\vert>\kappa} \frac{d^{3}p \ d^{3}q}{(2\pi)^{3}} \ e^{i\textbf{p}\cdot\textbf{x}}e^{-i\textbf{q}\cdot\textbf{y}} \ \langle 0\vert \phi^{B}_{\textbf{p}}(x_{0})(\phi^{B}_{\textbf{q}}(y_{0}))^{\ast}\vert 0\rangle=   \\
&=\int \frac{d^{3}p }{(2\pi)^{3}} \ e^{i\textbf{p}\cdot(\textbf{x}-\textbf{y})}\bigg(-\frac{2}{2\pi i}\int_{\Omega_{B}}^{+\infty}\frac{dp_{0} \ \kappa}{\sqrt{p_{0}^{2}-\Omega_{B}^{2}}} \ \frac{e^{-ip_{0}(x_{0}-y_{0})}}{p_{0}^{2}-\omega_{\textbf{p}}^{2}}\bigg) \ , \\
\langle 0\vert (\phi^{B}(y))^{\ast}\phi^{B}(x)\vert 0\rangle&=\int_{\vert\textbf{p}\vert,\vert\textbf{q}\vert>\kappa} \frac{d^{3}p \ d^{3}q}{(2\pi)^{3}} \ e^{-i\textbf{p}\cdot\textbf{x}}e^{i\textbf{q}\cdot\textbf{y}} \ \langle 0\vert (\phi^{B}_{\textbf{q}}(y_{0}))^{\ast}\phi^{B}_{\textbf{p}}(x_{0})\vert 0\rangle=  \\
 &=\int \frac{d^{3}p }{(2\pi)^{3}} \ e^{-i\textbf{p}\cdot(\textbf{x}-\textbf{y})}\bigg(-\frac{2}{2\pi i}\int_{\Omega_{B}}^{+\infty}\frac{dp_{0} \ \kappa}{\sqrt{p_{0}^{2}-\Omega_{B}^{2}}} \ \frac{e^{ip_{0}(x_{0}-y_{0})}}{p_{0}^{2}-\omega_{\textbf{p}}^{2}}\bigg) \ . \end{split} 
\end{equation}
The $ p_{0} $ integrals in this last expressions are ill defined since the poles $ \pm\omega_{\textbf{p}} $ are on the path of integration ($ \omega_{\textbf{p}}>\Omega_{B} $), however, we can adopt the prescription in the spectral function (\ref{spectral function}) to avoid the poles and thus the vacuum expectation value of half the anti-commutator of the fields becomes
\begin{equation}\label{correlation B}
\frac{1}{2}\langle 0\vert \lbrace\phi^{B}(x),(\phi^{B}(y))^{\ast}\rbrace\vert 0\rangle=-\frac{1}{2\pi i}\int_{\vert\textbf{p}\vert>\kappa} \frac{d^{3}p}{(2\pi)^{3}} \ e^{i\textbf{p}\cdot(\textbf{x}-\textbf{y})}\int_{\Omega_{B}}^{+\infty}\frac{dp_{0} \ \kappa}{\sqrt{p_{0}^{2}-\Omega_{B}^{2}}} \ \frac{e^{-ip_{0}(x_{0}-y_{0})}+e^{ip_{0}(x_{0}-y_{0})}}{p_{0}^{2}-\omega_{\textbf{p}}^{2}+i\varepsilon} \ . 
\end{equation}
In this equation the $ p_{0} $ integral is just the integral $ \mathcal{I}_{\textbf{p}}^{B} $ defined in (\ref{integral IB}). We thus obtain that the Hadamard function of the non-local field is equal to the restriction $ i\Delta^{\kappa}_{F}(x-y)\big\vert_{B} $ of the $ \kappa $-deformed Feynman propagator to trans-Planckian momenta
\begin{equation}\label{combination wightman con I}
\frac{1}{2}\langle 0\vert \lbrace\phi^{B}(x),(\phi^{B}(y))^{\ast}\rbrace\vert 0\rangle=i\int_{\vert\textbf{p}\vert>\kappa} \dfrac{d^{3}p }{(2\pi)^{4}} e^{i\textbf{p}\cdot(\textbf{x}-\textbf{y})} \ \mathcal{I}_{\textbf{p}}^{B}(x_{0}-y_{0})\equiv i\Delta^{\kappa}_{F}(x-y)\big\vert_{B} \ .
\end{equation}

Considering now the relations (\ref{kappa deformed feynman expand 2}), (\ref{t-ordered A con I}) and (\ref{combination wightman con I}), we can finally write down the expression of $i\Delta^{\kappa}_{F}(x-y)  $ in terms of the vacuum expectation values of the non-local scalar fields on ordinary Minkowski space.  The $ \kappa $-deformed Feynman propagator coincides with the T-product of such non-local fields for sub-Planckian momenta and to half their Hadamard function for trans-Planckian momenta
\begin{equation}\label{final relation propagator two-point}
i\Delta^{\kappa}_{F}(x-y)=\langle 0\vert T\big\lbrace\phi^{A}(x)(\phi^{A}(y))^{\ast}\big\rbrace\vert 0\rangle+\frac{1}{2}\langle 0\vert \lbrace\phi^{B}(x),(\phi^{B}(y))^{\ast}\rbrace\vert 0\rangle \ .
\end{equation}
Let us note that in the limit $ \kappa\rightarrow\infty $ the r.h.s. of this equation approaches the time-ordered two-point function as a consequence of the fact that in this limit the region $ B $ disappears.
 
Although the relation (\ref{final relation propagator two-point}) may look unusual, since it contains a splitting that depends on the value of the spatial momenta, it turns out that a similar expression for the propagator emerges in an apparently unrelated context, namely for the Feynman propagator of a degenerate Fermi gas (see e.g. the standard textbook reference \cite{Greiner}). In this case the ``deformation scale" is set by the Fermi momentum $k_{F} $, and the standard Feynman propagator for the Dirac field is recovered in the limit $ k_{F}\rightarrow 0 $, so that, in particular, the Fermi momentum plays the role of an IR deformation parameter. Indeed, when one replaces the standard vacuum with a noninteracting Fermi gas of electrons at zero temperature with Fermi momentum $ k_{F} $, the resulting propagator coincides with the advanced propagator for spatial momenta $ \textbf{p} $ lower than $ k_{F} $ and to the ordinary time-ordered two-point function for $ \vert\textbf{p}\vert> k_{F} $. A physical interpretation of such modification can be given by recalling that the standard Feynman propagator corresponds to an advanced propagation of the negative-energy solutions and to a retarded propagation of the positive-energy solutions. Now, in a degenerate Fermi gas, all the levels in the positive-energy electron continuum are occupied up to the Fermi momentum $ k_{F} $. Excitations related to these energy levels have to be treated like negative-energy states, i.e. they propagate backwards in time via the advanced propagator. Accordingly, when we consider states with an energy below the corresponding Fermi energy $ E_{F}=\sqrt{k_{F}^{2}+m^{2}_{e}} $, both the negative-energy solutions and the positive-energy solutions propagates in an advanced way, conversely, when $ \vert\textbf{p}\vert>k_{F} $, negative-energy solutions still propagate in an advanced way while the positive-energy solutions propagate in a retarded way, so that we recover the standard structure that gives rise to the time-ordered two-point function.
  
 In the analysis we presented the deformation parameter $ \kappa $, usually identified with the Planck energy $ E_{p} $, is instead a UV scale and the $ \kappa $-deformed Feynman propagator differs from the time-ordered two-point function for spatial momenta bigger than $ \kappa $. Indeed, for trans-Planckian momenta the $\kappa$-deformed propagator has the form of the Hadamard function which, being the vacuum expectation value of the anti-commutator of the field, does not posses a time ordering. Nonetheless, like the T-product, the anti-commutator is symmetric under exchange of the spacetime arguments of the fields. A physical interpretation for the appearance of the Hadamard function for trans-Planckian momenta is however less straightforward than the interpretation of the propagator structure of a degenerate Fermi gas. Indeed, we now deal with vacuum expectation value of a non-local field, in which both the simple poles and the branch cuts contributes. The expression (\ref{final relation propagator two-point}) for the $ \kappa $-deformed Feynman propagator, suggests that for trans-Planckian momenta we lose the notion of time orientation; the two Wightman functions $ \langle 0\vert \phi^{B}(x)(\phi^{B}(y))^{\ast}\vert 0\rangle $ and $ \langle 0\vert (\phi^{B}(y))^{\ast}\phi^{B}(x)\vert 0\rangle $ contribute both for $ x_{0}>y_{0} $ and $ x_{0}<y_{0} $.

\section{Discussion}
In this work we provided a comprehensive description of the structure and properties of the Feynman propagator of a $\kappa$-deformed field theory. We started from its derivation from a non-commutative generating functional and described the non-trivial singularity structure determined by the curved geometry of momentum space dual to the $\kappa$-Minkowski space. We showed how such singularity structure is responsible for the new features of the $\kappa$-Feynman propagator, and how such features are intimately related to the different behaviour of the propagator for sub and trans-Planckian field modes. Our results showed that the propagation of perturbations in $\kappa$-deformed field theory can be equivalently described in terms of perturbations generated by a {\it fuzzy} source, which cannot be sharply localized in spacetime, or by a ordinary source which generates perturbations mediated by a $\kappa$-deformed propagator whose non-trivial spacetime profile was analyzed in detail.

Our analysis also addressed the question of whether the $\kappa$-deformed Feynman propagator derived from the generating functional, which we showed to be a Green's function of the $\kappa$-deformed Klein-Gordon operator, can be related to the vacuum expectation value of suitable products of field operators. In order to explore such relationship we resorted to a mapping of non-commutative fields on $\kappa$-Minkowski space to fields on ordinary Minkowski space with a non-local kinetic term. Adopting techniques previously developed in the literature for the canonical quantization of non-local field theories, we showed that the $\kappa$-deformed Feynman propagator can be related to the vacuum expectation value of products of non-local field operators. We obtained that for sub-Planckian field modes the propagator coincides with the vacuum expectation value of the time-ordered product of non-local fields, while for trans-Planckian momenta it is related to the non-local Hadamard function, i.e. the vacuum expectation value of the {\it anti-commutator} of non-local field operators. This peculiar behaviour is reminiscent of the propagation of electrons in a Fermi gas where above and below the Fermi momentum their propagator is described by different kinds of two-point functions. While electrons above the Fermi momentum propagate as ordinary free electrons, excitations below the Fermi level (holes) behave has antiparticles and are described by an advanced propagator. In our case the deformation parameter $\kappa$ sets a UV momentum scale above which propagation of excitations is described by the Hadamard propagator in which a notion of time ordering is no longer present. This result provides an interesting picture of how the non-commutativity of $\kappa$-Minkowski space affects the propagation of particles in the deep UV and provides valuable insight for understanding the physical properties of this particular non-commutative deformation of quantum field theory.

While the main focus of our analysis has been on the study of $\kappa$-deformed propagation, it would be interesting to further explore the alternative picture of $\kappa$-deformation as ordinary propagation of perturbations from a smeared out source as discussed in Section 3.2. In particular it should be in principle possible to reformulate this scenario in terms of an equivalent non-local field theory on ordinary Minkowski spacetime and investigate its non-trivial UV effects. We postpone such study to future work.

%\section*{Acknowledgements}

\end{document}